\documentclass[11pt]{article}
\usepackage[margin=1.85cm]{geometry}
\usepackage[utf8]{inputenc}
\usepackage[backref=page]{hyperref}
\usepackage[round]{natbib}
\usepackage[dvipsnames]{xcolor}
\usepackage{amsmath,amsthm,amssymb,nicefrac}
\usepackage{tikz, algorithm, algpseudocode}
\usepackage{csquotes, booktabs, url, float, caption}
\def\EE{{\mathbb E}}    
\def\11{{\mathbf 1}}    

\def\cN{{\mathcal N}}

\def\d{\,{\mathrm d}}

\title{Score-based generative emulation of \\ impact-relevant Earth system model outputs}

\author{Shahine Bouabid, Andre Nogueira Souza, Raffaele Ferrari \\
\small\textit{Department of Earth, Atmospheric, and Planetary Sciences} \\ 
\small\textit{Massachusetts Institute of Technology, Cambridge, MA, United States}}
\date{}

\begin{document}
\maketitle

\begin{abstract}

Policy targets evolve faster than the Coupled Model Intercomparison Project cycles, complicating adaptation and mitigation planning that must often contend with outdated projections. Climate model output emulators address this gap by offering inexpensive surrogates that can rapidly explore alternative futures while staying close to Earth System Model (ESM) behavior. The focus is on emulators designed to provide inputs to impact models. Using monthly ESM fields of near-surface temperature, precipitation, relative humidity, and wind speed, it is shown that deep generative models have the potential to model the joint distribution of variables relevant for impacts. The specific model proposed uses score-based diffusion on a spherical mesh and runs on a single mid-range graphical processing unit. A thorough suite of diagnostics is introduced to compare emulator outputs with their parent ESMs, including their probability densities, cross-variable correlations, time of emergence, or tail behavior. The emulator performance is evaluated across three distinct ESMs in both pre-industrial and forced regimes. The results show that the emulator produces distributions that closely match the ESM outputs and captures key forced responses. They also reveal important failure cases, notably for variables with a strong regime shift in the seasonal cycle. Although not a perfect match to the ESM, the inaccuracies of the emulator are small relative to the magnitude of internal variability in ESM projections. This suggests that the generative emulators can be useful in supporting impact assessment. Priorities for future development toward daily resolution, finer spatial scales, and bias-aware training are discussed.

\end{abstract}

\section{Introduction}\label{section:introduction}

Planning adaptation measures and evaluating mitigation choices depends on assessing the projected impacts of climate change under different future scenarios. The computational cost of Earth System Models (ESMs) has prevented them from keeping pace with updates to these scenarios. For example, although the set of scenarios for the Phase 7 Coupled Model Intercomparison Project (CMIP) is expected to be released this year~\citep{van2025scenario}, the corresponding simulations will not be available for another few years. Unless the handful of off-the-shelf scenarios already run by ESMs are deemed sufficient to cover future alternative pathways of emissions, fast ESM surrogates, called \emph{climate model output emulators} can be extremely valuable~\citep{jones2024bringing, tebaldi2025emulators}. Typically, these emulators do not attempt to reproduce the dynamics of the climate system. Rather, for a subset of variables relevant in applications like impact modeling, they aim to reproduce statistics of the \emph{distribution of climate model outputs}. The goal is to learn statistical features that best characterize the distribution produced by an ensemble of simulations from the reference ESM. This is a crucial nuance, since under this rationale, an emulator only needs to provide a computationally efficient generator that retains the same statistical behavior as the original ESM.

The Inter-Sectoral Impact Model Intercomparison Project (ISIMIP)~\citep{warszawski2014inter} is a concerted effort to collect, harmonize, and distribute data relevant for impact assessment across sectors and future scenarios. It presents a selection of \href{https://protocol.isimip.org/#/ISIMIP3a/32-climate-related-forcing}{11 atmospheric climate model output variables} that are most consequential for sectors such as agriculture, energy demand, or forestry. This selection defines a priority list for any emulator intended to support physical-risk assessment. The past decade has seen a profusion of emulator designs which, collectively, have demonstrated skillful reproduction of these variables at the regional scale, with orders-of-magnitude lower computational cost than running new ESM experiments. They include emulators that target selected low-order statistics of climate model output variables such as traditional pattern scaling~\citep[e.g.][]{santer1990developing, tebaldi2014pattern, herger2015improved} and impulse-response models~\citep[e.g.][]{lucarini2017predicting, freese2024spatially, womack2025rapid, winkler2025towards, sandstad2025meteorv1}, but also emulators targeting the full probability distribution~\citep[e.g.][]{beusch2020emulating, nath2022mesmer, geogdzhayev2025eof, mathison2025rapid}. In theory, this last category supersedes emulating selected statistics, since any statistic can be derived from the distribution.

A survey of the existing \enquote{\emph{anthropogenic forcing $\to$ climate}} emulators shows that most focus on just one or two variables at a time~\citep[e.g.][]{snyder2019joint, nath2024mercury, schongart2024introducing,tebaldi2025emulators}, often making parametric assumptions for the variables distributions based on generalized linear models or Gaussian processes~\citep[e.g.][]{castruccio2014statistical, link2019fldgen, goodwin2020computationally, quilcaille2022showcasing, bouabid2024fairgp}. This has proven to be sufficient, especially when the variables of interest are sufficiently aggregated (spatially or temporally) to be well captured by parametric forms; for example, a lognormal distribution for precipitation or a Gaussian one for temperature. However, these approaches become prone to misspecification and are increasingly cumbersome to implement when the objective is to emulate dozens of variables jointly at high resolution, which is needed to assess high-impact climate damages that are often associated with compound risks due to co-occurring events~\citep{zscheischler2020typology, mathison2023description}. Capturing joint dependencies requires emulating not only individual variables, but also spatio-temporal and cross-variable correlations, and the number of correlations grows quadratically with the number of variables and spatial resolution.

Alternative strategies for the coherent emulation of multiple variables often rely on some flavour of nearest-neighbor matching between the emulated scenario and existing ESM simulations. These include \enquote{stitching} strategies, which connect time slices of existing ESM runs, and can provide most atmospheric variables jointly~\citep{tebaldi2022stitches, byers2025fast}. Another example is the approach of \citet{kitsios2023machine}, which uses dimensionality reduction to emulate the forced response of an arbitrary set of variables, and draws internal variability from the closest available ESM realization. While effective in preserving multivariate consistency, such methods rely on having large volumes of ESM output archives at hand, and by construction cannot generate new realizations of ESM variability.

Deep generative models offer an appealing alternative: they learn the full joint distribution directly, are well-suited to high-dimensional structured data, and provide a compressed representation of ESM outputs. Recent studies have already demonstrated their ability to emulate complete atmospheric states on multiple pressure levels when forced by a prescribed sea surface temperature~\citep{watt2023ace, brenowitz2025climate}. While the design of these emulators is primarily focused on accelerating atmospheric simulation for scientific discovery, their success motivates extending them to the impact-focused setting considered here.

In this work, we propose a score-based diffusion emulator that learns the high‑dimensional joint probability distribution of monthly climate model fields conditioned on the global mean surface temperature (GMST) anomaly. The emulator produces data on the ESM’s native grid while carrying out its computations on an equal-area spherical HEALPix mesh similar to the setup of \mbox{\citet{brenowitz2025climate}}. A pattern scaling step maps GMST onto regional mean temperature anomalies, providing spatial structure as conditioning information to the emulator. The proposed model is purposefully designed to be lightweight and run efficiently on a single mid-range Graphical Processing Unit (GPU). We evaluate the model on three distinct ESMs and for a selection of four surface variables relevant for impact assessment: surface temperature, precipitation, relative humidity, and wind speed. Although we restrict ourselves to these four fields, the emulator can accommodate additional variables without modifications to its design. We find that the emulator learns to sample from the joint distribution of these variables at the climate model's native spatial resolution. It reproduces internal variability and response to forcing. While many impact studies require daily or finer resolution, the current emulator is already useful for applications relying on long-term averages, and provides a foundation for higher-resolution approaches as well as for pairing with downscaling techniques.

The paper is organized as follows. Section~\ref{section:data&methods} describes the data, the emulator design, and introduces evaluation metrics. Section~\ref{section:unforced} presents the results for the emulation of unforced simulations, and Section~\ref{section:forced} evaluates the forced response of the emulator. Section~\ref{section:discussion} discusses limitations and outlooks and Section~\ref{section:conclusion} concludes.

\section{Data and Methods}\label{section:data&methods}

Section~\ref{subsection:data} describes the CMIP6 simulations used in training and evaluation. Section~\ref{subsection:setup} sets the notations and formal objective of emulation. Sections~\ref{subsection:pattern-scaling}, \ref{subsection:diffusion}, and \ref{subsection:architecture} outline, respectively, the use of pattern scaling as a conditioning field, the score-based diffusion framework, and the neural network architecture. Finally, Section~\ref{subsection:distances} introduces the metrics used to assess emulator performance.

\subsection{Data}\label{subsection:data}

We select variables informed by climate forcing priorities from the ISIMIP protocol, starting with 2m air temperature, precipitation, 2m relative humidity, and 10m wind speed. Other variables (e.g., specific humidity, daily maximum and minimum temperature, downwelling longwave radiation) can be incorporated without changes to the emulator design, but we choose in this work to focus on the four listed above for clarity of presentation. We retain the native spatial resolution of the data ($\sim$1-2$^\circ$) and use monthly averages to keep data volumes manageable, although the same approach can extend to daily data. Each variable is pre-processed to be an anomaly relative to the climatology of its model's pre-industrial control run (piControl); unless stated otherwise, every variable in this work represents a positive or negative departure from that baseline.

\begin{figure}[htbp]
    \centering
    \includegraphics[width=0.5\linewidth]{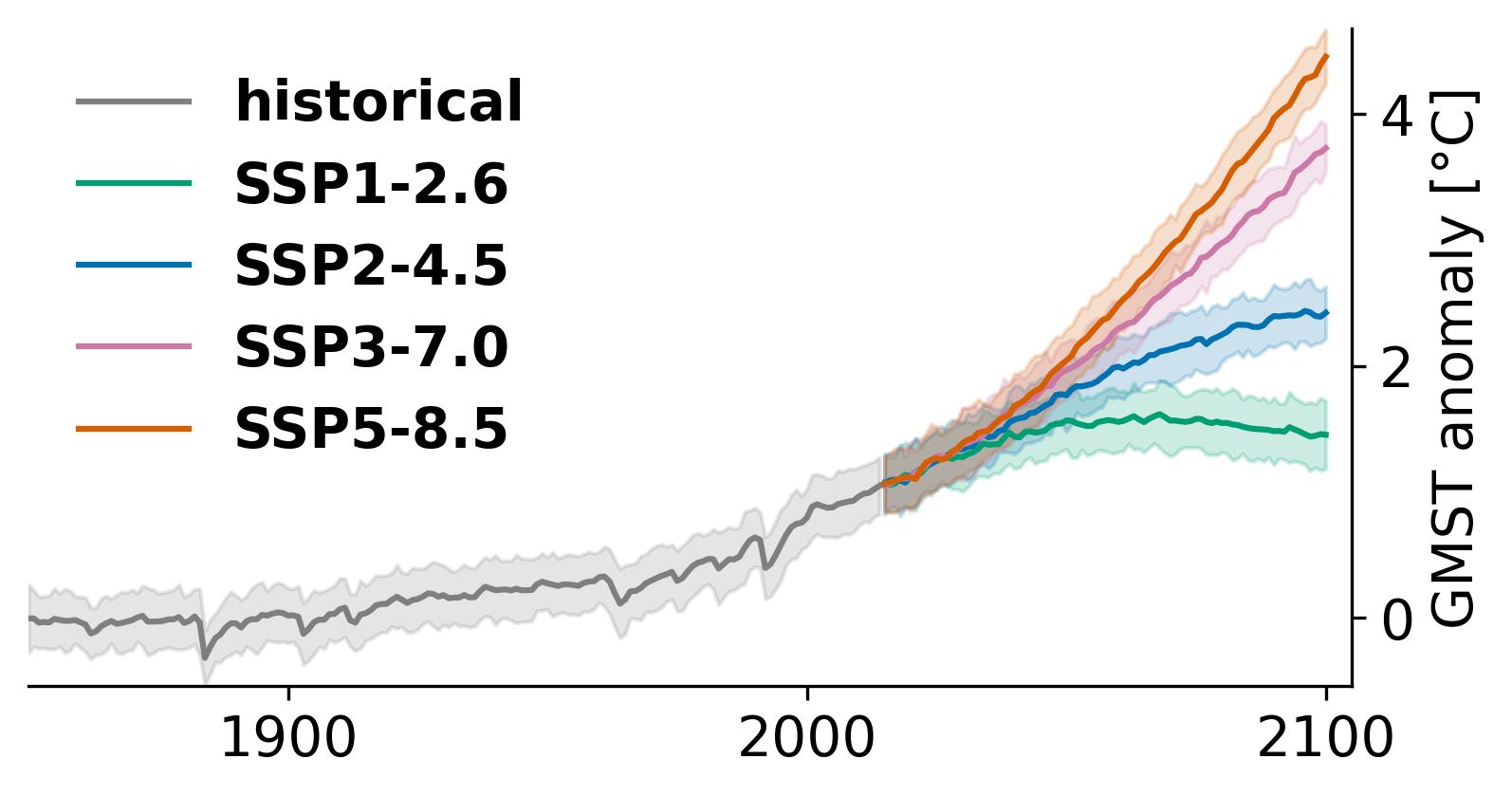}
    \caption{Time series of annual GMST anomalies from the MPI-ESM2-1-LR ScenarioMIP simulations~\citep{mpidatascenariomip}. Anomalies are relative to the climatology of the model's piControl run. The solid line shows the ensemble mean across 50 members, and shading indicates two standard deviations.}
    \label{fig:ssps}
\end{figure}

Our focus is on climate model runs over scenarios from the CMIP6 ScenarioMIP protocol~\citep{o2016scenario} since they constitute the most widely used projections for impact assessment. The associated Shared Socioeconomic Pathways (SSPs)~\citep{riahi2017shared} shown in Figure~\ref{fig:ssps}, specify different 2014-2100 trajectories for long- and short-lived emissions. Each pathway produces a distinct global mean surface temperature (GMST) trajectory, ranging from Paris-aligned to strongly overshooting. To train the emulator, we use ESM outputs from SSP1-2.6 and SSP5-8.5, and complement these with historical and piControl experiments from the core CMIP6 protocol~\citep{eyring2016overview}. This constitutes a diverse training set of different emulation regimes, including unforced outputs, strong curbing of emissions, and an increase in emissions. When presenting experiments, we will explicitly mention the scenarios used to evaluate the emulator, as they can vary for different diagnostics.

A large ensemble is necessary for the generative model to learn to disentangle the model's internal variability from the anthropogenic signal~\citep{lutjens2024impact}. We use large-ensemble simulations from three CMIP6 models --- MPI-ESM1-2-LR, MIROC6, and ACCESS-ESM1-5 --- each with over 40 members covering historical and SSP scenarios~\citep{mpidatacmip, mpidatascenariomip, mirocdata, mirocdatascenariomip, accessdata, accessdatascenariomip}. Each emulator is trained on data from a single ESM; data from multiple models supports inter-model and emulator comparison, not cross-model training. Table C.1 in Section C of the supplementary summarizes the data used, including simulation length, ensemble size, and train-test split.

We draw the reader’s attention to the fact that in the main text, we chose to primarily present results from emulating the MPI-ESM1-2-LR model. Since many of the same conclusions about the emulator’s skill hold across all three ESMs, there is limited value in reproducing every diagnostic in full. We therefore focus on one model to highlight the methodological insights that are largely invariant to the choice of ESM. However, complete figures and diagnostics for all three models are provided in the supplementary material, and inter-model differences are discussed in the main text where relevant.

\subsection{Problem setup}\label{subsection:setup}

Let $p_{ESM}$ denote the ensemble distribution produced by realizations of a reference ESM given a forcing scenario; that is, consider that individual runs from the ESM correspond to samples drawn from this distribution. We denote $x_t$ a monthly averaged climate model output anomaly at time $t$, where $t$ indexes a year and a calendar month, of the climate variables $x$. It can be thought of as a state, relative to the pre-industrial control climatology, that the climate model could realize at time $t$, given a forcing trajectory, 
\begin{equation}
    x_t \sim p_{ESM}(x \mid \text{Forcing}_{\leq t}).
\end{equation}
In technical terms, $p_{ESM}$ represents the distribution of the forward model induced by the ESM for slightly perturbed initial conditions, given a prescribed forcing. Since our focus is on the development of emulators useful for impact assessment, for practical purposes, $x_t$ can be thought of as multiple concatenated spatial maps of surface variables selected by their relevance for risk assessment.

Because regional climate outcomes tend to display a significant correlation with the GMST anomaly $\Delta T_t$, numerous existing climate model output emulators have adopted it as their main driver~\citep{nath2022mesmer, quilcaille2023extending, schongart2024introducing, nath2024mercury, geogdzhayev2025eof, wang2024stochastic}. This dependency is convenient because the GMST response to changes in emissions can be easily derived from simple climate models --- which offer a reliable lower-order approximation of the GMST response to long- and short-lived forcers~\citep{meinshausen2011emulating, gasser2017compact, leach2021fair} --- or directly from integrated assessment models. This makes emulators ready for the full \emph{emissions $\to$ regional climate} pipeline. Formally, the underlying assumption in this shared design for many emulators is that
\begin{equation}
    p_{ESM}(x \mid \text{Forcing}_{\leq t}) \approx p_{ESM}(x \mid \Delta T_t).
\end{equation}

This is, of course, an oversimplifying assumption, which dismisses the effect of forcers' regionality, the different types of forcing agent, or the climate system memory. Nonetheless, evidence from emulator development literature suggests that replacing the forcing with $\Delta T_t$ already captures a significant fraction of projected variability under plausible future scenarios. This is also expected to align with the requirements of the upcoming FASTMIP emulator intercomparison exercise~\citep{seneviratne2024using}. We therefore proceed with this assumption in the remainder of the manuscript and revisit its implications in the discussion. In this context, our objective can be framed as finding a surrogate conditional distribution $q(x \mid \Delta T_t)$, which we can easily sample from, and such that

\begin{equation}
    d\big(q, p_{ESM}\big) \leq \epsilon,
\end{equation}
where $d$ is a measure of the discrepancy between probability distributions and $\epsilon > 0$ is a selected threshold. The preferred notions of distance will depend on which statistical features are assessed (e.g., mean, variance, tails), and we discuss in Section~\ref{subsection:distances} different choices for $d$ that we use to evaluate the emulator.

\subsection{Pattern scaling conditioning}\label{subsection:pattern-scaling}

We introduce spatial information in the conditioning signal based on a linear relationship between the global and local mean surface temperature anomaly, known as \emph{pattern scaling}. Pattern scaling is an emulation technique that assumes the forced regional response of a climate variable scales linearly with an indicator of the global climate response~\citep{santer1990developing}. For regional surface temperature, such a fixed warming pattern consistently emerges across CMIP6 simulations as a function of GMST~\citep{tebaldi2014pattern, osborn2018performance} and is supported by energy balance arguments for medium-to-high emission scenarios~\citep{giani2025origin}.

While other surface variables tend to conform less to this linear approximation~\citep{kravitz2017exploring}, it is recognized that pattern scaling captures the slow large-scale regional changes simulated by ESMs, which matter on a centennial time scale~\citep{tebaldi2014pattern, lee2021future}. Therein, owing to its simplicity and competitiveness in benchmarks~\citep{watsonparris2022climatebench, lutjens2024impact}, pattern scaling remains today a recurrent component in the design of emulators aiming to support impact assessment questions~\citep{osborn2016pattern, link2019fldgen, beusch2020emulating, tebaldi2020emulating, nath2022mesmer, quilcaille2022showcasing, quilcaille2023extending, schongart2024introducing, mathison2025rapid}.

Pattern scaling does not represent internal variability and, therefore, cannot serve as an emulator of the climate model output distributions. However, it offers an inexpensive way to augment the conditioning signal with spatial structure. We leverage this and condition our emulator on the projected pattern scaling for regional temperatures rather than $\Delta T_t$. Using the training data, we fit a separate pattern scaling model for each month $P^{\texttt{month}}_{t} = \beta^{\texttt{month}}_1\Delta T_t + \beta^{\texttt{month}}_0$, where $\beta^{\texttt{month}}_1, \beta^{\texttt{month}}_0$ are fixed spatial patterns of warming for a given month. At each time step $t$, the scalar $\Delta T_t$ is thus mapped to a spatially resolved projection $P^{\texttt{month}}_t$ which is passed to the emulator, providing it with structured information about the forced regional response in surface temperature.

In the rest of the paper, we continue to write the conditioning variable as $\Delta T_t$, interpreting pattern scaling as a design step in constructing $q$. This emphasizes that GMST remains the fundamental driver: to emulate a new scenario, one provides the emulator with the associated GMST trajectory, which is then mapped through a pattern scaling model fitted to the training data.

\subsection{Score-based diffusion emulator}\label{subsection:diffusion}

We model $q$ using a score-based diffusion framework~\citep{ho2020denoising, song2020score}, motivated by its suitability for high-dimensional structured data and its demonstrated effectiveness in climate data generation~\citep{bassetti2024diffesm, mardani2025residual, hess2025fast, brenowitz2025climate}. However, we emphasize that alternative approaches to learn a surrogate generative process to the climate model --- such as variational inference, particle filters, or flow-based generative models --- could also be considered and may be better suited to certain applications.

Score-based diffusion models rely on gradually transforming samples from a simple Gaussian distribution into samples from a complex distribution of interest, in our case $p_{ESM}$. Learning this transformation involves two steps: first, $p_{ESM}$ is gradually mollified until it becomes indistinguishable from a Gaussian distribution; second, a neural network is trained to reverse this process. In practice, we carry out the first step by adding Gaussian noise to climate model outputs $\tilde x_t = x_t + \varepsilon$, where $\varepsilon\sim \cN(0, \sigma^2)$ with variance $\sigma^2$. This can be viewed as sampling from a \enquote{smoothed} version of the climate model output distribution defined by
\begin{equation}
    p_{ESM}^\sigma = p_{ESM} * \cN(0, \sigma^2),
\end{equation}
where $*$ denotes convolution. By repeating the convolution with increasing variance levels, the distribution becomes progressively smoother, and ultimately is indistinguishable from a Gaussian distribution for a sufficiently large $\sigma^2$. Then, we train a neural network $f_\theta$ to reverse this transformation. We use least-squares regression, predicting the original climate model output from its transformed Gaussian versions at different variance levels $\sigma$, by minimizing the loss
\begin{equation}
    \EE\big\|x_t - f_\theta(\tilde x_t, \sigma, \Delta T_t)\big\|^2.
\end{equation}

Such direct regression will generally fail to recover the original climate model output, especially when the injected Gaussian noise is large. However, the key to the success of score-based generative modeling is that the neural network $f_\theta$ does not attempt to reverse the transformation outright. Rather, it identifies a vector that nudges transformed samples toward regions of higher density in the climate model output distribution. This vector is known as the \emph{score function} of the smoothed climate model output distribution, and is defined as the gradient of its log-density $\nabla \log p^\sigma_{ESM}$. Tweedie's formula~\citep{efron2011tweedie} provides a useful explanation: regressing away Gaussian noise is mathematically equivalent to moving the transformed sample in the direction of the score,
\begin{equation}~\label{eq:tweedie}
	\EE[x_t\mid \tilde x_t, \sigma, \Delta T_t] = \tilde x_t + \sigma^2 \nabla \log p^\sigma_{ESM}(\tilde x_t \mid \Delta T_t).
\end{equation}
By the property of least-square regression, the trained neural network $f_\theta$ constitutes an estimator of the conditional expectation in Eq.~\ref{eq:tweedie}, and allows for estimating the score function:
\begin{equation}
    \nabla \log q^\sigma(\tilde x_t \mid \Delta T_t) = \frac{f_\theta(\tilde x_t, \sigma, \Delta T_t) - \tilde x_t}{\sigma^2},
\end{equation}
where $q^\sigma$ denotes our learned approximation of the smoothed distribution. This score estimate provides a direction in which we can iteratively displace the noised samples to reverse the transformation process.

This allows us to emulate the climate model output distribution by starting from a new Gaussian sample $\tilde x\sim \cN(0, \sigma_{\max}^2)$ for a sufficiently large $\sigma^2_{\max}$, and evolving it in the direction of the score for a decreasing $\sigma$ to produce a sample from the estimated climate model output distribution. The emulated climate model output distribution $q$ is then defined as $q^{\sigma}$ for $\sigma\to 0$, and this offers is a procedure to draw samples from $q$. The choice of $\sigma_{\max}^2$, schedule for noise levels, and the integration scheme used to follow the direction of the score are important design parameters of this sampling algorithm. We detail our technical choices in Section A of the supplementary.

\subsection{HEALPix UNet architecture}\label{subsection:architecture}

However compelling the score-based diffusion framework can be, diffusion models in reality owe their effectiveness to advances in neural network architectures for structured data, and the computational solutions to support them. For spatially structured data on a square, a common choice of architecture for $f_\theta$ is the UNet~\citep{ronneberger2015u}. Its effectiveness can be understood through its multi-resolution structure, which implicitly performs transformations analogous to processing signals in a hierarchical wavelet basis, separating information at different spatial scales~\citep{williams2023unified}. This, however, does not support data on a sphere.

To respect Earth's geometry, we implement a variation of the classical UNet architecture that operates on an equal-area HEALPix mesh~\citep{gorski2005healpix}. This mesh first splits the domain into 12 base diamonds that are exactly equal in area. Each refinement level subdivides every diamond into 4 self-similar diamonds, so the equal-area property is preserved. This avoids the pole-to-equator cell-size distortion and longitude-wrapping discontinuities that arise in equiangular grids, while preserving a hierarchical neighborhood structure ideal for multi-resolution processing. This choice of mesh was also made by \citet{karlbauer2024advancing} and \citet{brenowitz2025climate}, and therefore our architectures bear strong similarities.

We expect that impact assessment workflows will generally assume climate model outputs on a regular lat-lon grid, and therefore keep the ESM’s native gridding for inputs and outputs. A pair of lightweight bipartite graph neural networks maps fields from the equiangular grid to HEALPix for processing and back again, allowing the model to satisfy spherical fidelity internally while being compatible with climate model output data on the ESM's native grid. Technical details on the neural network design choices and its training procedure are provided in Section B and C of the supplementary.

To ensure the emulator is accessible to a wide range of stakeholders, we designed the architecture to be lightweight. By focusing on a small set of variables relevant for impact assessment, the network is kept compact, with approximately 10 million parameters. This number of parameters is an order of magnitude smaller than cBottle~\citep{brenowitz2025climate}, a score-based diffusion model developed to emulate global, kilometer-scale climate simulations. cBottle differs from our emulator by using sea surface temperature as a conditioning input to emulate entire atmospheric atmospheric states comprising 44 variables, and therefore naturally requires more capacity. The total file size of our emulator is 50 MB and can be downloaded from our repository. This lightweight design allows the emulator to run on a single mid-range GPU; generating one sample takes about 1 second on a T4 (mid-range) and about 0.13 seconds on an H100 (state-of-the-art high-end). Because samples are generated through independent processes, this emulation can be parallelized to accelerate large ensemble generation.

\subsection{Distances between distributions}\label{subsection:distances}

Our goal is for the emulator distribution $q$ to closely approximate the ESM output distribution $p_{ESM}$ under a chosen notion of distance $d(q, p_{ESM})$. We describe below the notions of distances that we use to diagnose different aspects of the discrepancy between the distributions.

\paragraph*{Distance in individual statistics}

Throughout this work, we often define $d$ in terms of error in individual statistics --- such as means, variances, correlations, quantiles --- as it provides important insights on how two distributions can differ. Biases in moments show errors in the distribution's location, spread, or skewness; biases in correlations reveal mismatches in dependence structures; biases in quantiles, often more robust to outliers than moments, are particularly useful to assess differences in the distribution tails. In practice, comparing the actual statistics directly can often make interpretations more natural than comparing errors. Visual inspection can often be sufficient to judge whether emulator and ESM statistics align, and we therefore rely on it for multiple diagnostics in the following sections.

When reporting explicit distances is useful, we privilege defining $d$ in terms of error relative to a reference quantity. This makes the reported values easier to interpret and compare across variables. We use two strategies: relative errors and error-to-noise ratios. Relative errors express the error as a percentage of deviation from the ESM statistic, where an acceptable threshold is chosen by judgment (e.g.\ 10\%). The error-to-noise ratio is defined by dividing the error by the ESM’s internal variability. This is particularly relevant for evaluating errors in means: if the bias is smaller than the natural variability, it is unlikely to be detectable in individual realizations. This naturally sets an acceptable threshold of $d(q, p_{ESM}) \leq 1$, with larger values indicating a meaningful bias.

\paragraph*{Earth Mover Distance}

Beyond individual statistics, many useful metrics to compare probability distributions in their entirety exist, such as the Kullback-Leibler divergence, the total variation distance, or the maximum mean discrepancy. We use the Earth Mover Distance (EMD), also known as the 1-Wasserstein distance, because we find it to have a more intuitive interpretation when dealing with physical quantities. The EMD is computationally intractable for high-dimensional data. As such, alternatives like the maximum mean discrepancy are better suited as global summary metrics, at the expense of interpretability. This is particularly useful in machine learning, to guide model development. Here instead, we prioritize grid-cell level diagnostics to examine spatial patterns of error. In this setting, the comparison reduces to one-dimensional distributions, for which the EMD is tractable.

Imagine distributions as binned histograms of mass as a function of horizontal location. Then the work to displace a unit of mass against friction is proportional to its weight times the distance. The EMD between $p_{ESM}$ and $q$ corresponds to the minimal amount of work needed to rearrange the masses so that they exactly overlay. For a single grid cell $\texttt{gc}$ and variable $\texttt{var}$, the distributions are one-dimensional, and the optimal rearrangement simply pairs quantiles together~\citep{santambrogio2015optimal}. Thus, if $F_{ESM}$ and $ F_q$ are the cumulative distribution functions for the ESM and the emulator, the work needed to rearrange quantiles is given by
\begin{equation}\label{eq:emd}
    \operatorname{EMD}(\texttt{gc},\texttt{var}) = \int \Big|x_{\texttt{gc},\texttt{var}} - F_{ESM}^{-1}\circ F_{q}(x_{\texttt{gc},\texttt{var}})\Big|\; q(x_{\texttt{gc},\texttt{var}})\d x_{\texttt{gc},\texttt{var}}.
\end{equation}
The term $|x - F_{ESM}^{-1}\circ F_{q}(x)|$ represents the distance a unit of mass from $q$ must be shifted when $p_{ESM}$ is held fixed, and $q(x)\d x$ gives the amount of mass being displaced. From this expression, we see that the EMD is equally sensitive to small shifts of large bulks of probability or large shifts of small probability mass, since they represent the same amount of work.

\begin{figure}[h]
    \centering
    \includegraphics[width=\linewidth]{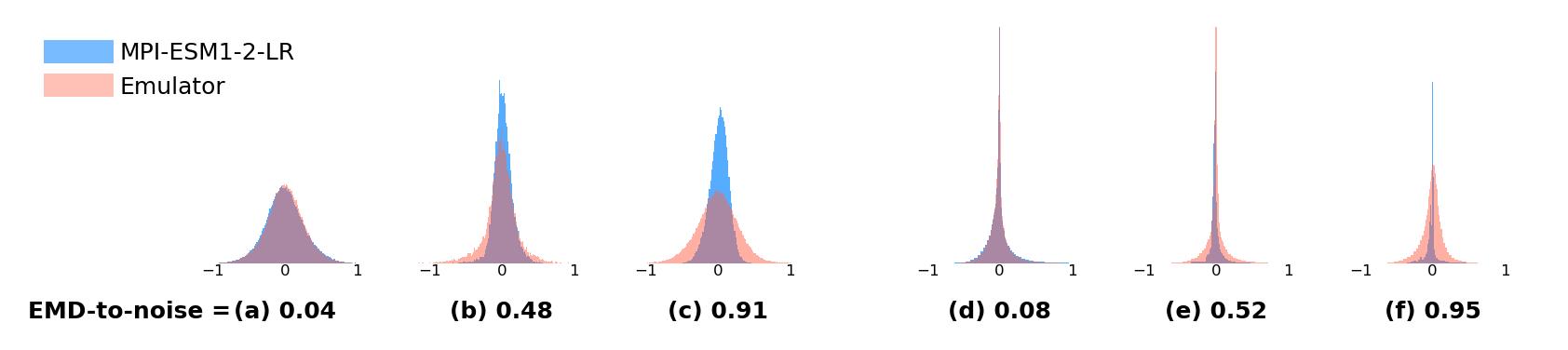}
    \caption{Distributions from the MPI-ESM1-2-LR piControl run and the emulator with $\Delta T_t = 0$, with their EMD-to-noise ratios. Distributions are normalized and shown on a common x-axis. Panels: (a) November wind speed, Mediterranean; (b) October relative humidity, western Africa; (c) July temperature, Arctic; (d) February precipitation, tropical Pacific; (e) July precipitation, Mediterranean; (f) December precipitation, central Africa.}
    \label{fig:emd-examples}
\end{figure}

Because probability masses are dimensionless and sum to one, Eq.~\ref{eq:emd} also implies that the EMD inherits the physical units and scale of the underlying variable, which makes it more easily interpretable. For example, if $p(T)$ is a distribution of temperatures, then the EMD to its shifted version $p(T +\delta T)$ exactly equals the shift in temperatures $|\delta T|$. Inspired by this property, we assess the significance of the misalignment between $q$ and $p_{ESM}$ by comparing their EMD to the typical internal variability at each grid cell. We introduce an EMD-to-noise ratio
\begin{equation}
    \operatorname{EMD-to-noise}(\texttt{gc}, \texttt{var}) = \frac{\operatorname{EMD}(\texttt{gc}, \texttt{var})}{\sigma_{\texttt{gc,var}}},
\end{equation}
where the noise $\sigma_{\texttt{gc,var}}$ is taken as the internal variability of the ESM. The ratio quantifies \textit{how many \enquote{standard deviation shifts equivalents} are required to realign the distributions.} Figure~\ref{fig:emd-examples} compares emulated distributions with their corresponding ESM outputs, together with the associated EMD-to-noise ratios. The apparent degree of misalignment between the distributions increases rapidly with the ratios. As a practical heuristic, we regard EMD-to-noise ratios below 0.5 as indicative that the emulator reproduces the ESM distribution within internal variability, whereas larger values are representative of a significant mismatch in distribution.

\section{Emulation of unforced simulations}\label{section:unforced}

We first evaluate if the emulator can represent the monthly internal variability of the climate model under stationary conditions without external forcing. Formally, this amounts to assessing whether in the absence of any forcing we have
\begin{equation}
    q(x\mid \Delta T_t = 0) \approx p_{ESM}(x \mid \text{Forcing}_{\leq t} = 0).
\end{equation}

To evaluate this, we train for each ESM an emulator on the piControl, historical, SSP1-2.6, and SSP5-8.5 runs. We then generate a large ensemble of emulated monthly samples in the absence of forcing and compare its statistics to the statistics of the climate model's piControl run. For a given month, we generate an ensemble of $1000$ samples by first drawing $\Delta T \sim \cN(0, \sigma_{PI}^2)$, where $\sigma_{PI}$ is the estimated GMST standard deviation from the piControl run, and then sampling from $q(x\mid \Delta T)$.

\begin{figure}[htbp]
    \centering
    \includegraphics[width=0.9\textwidth]{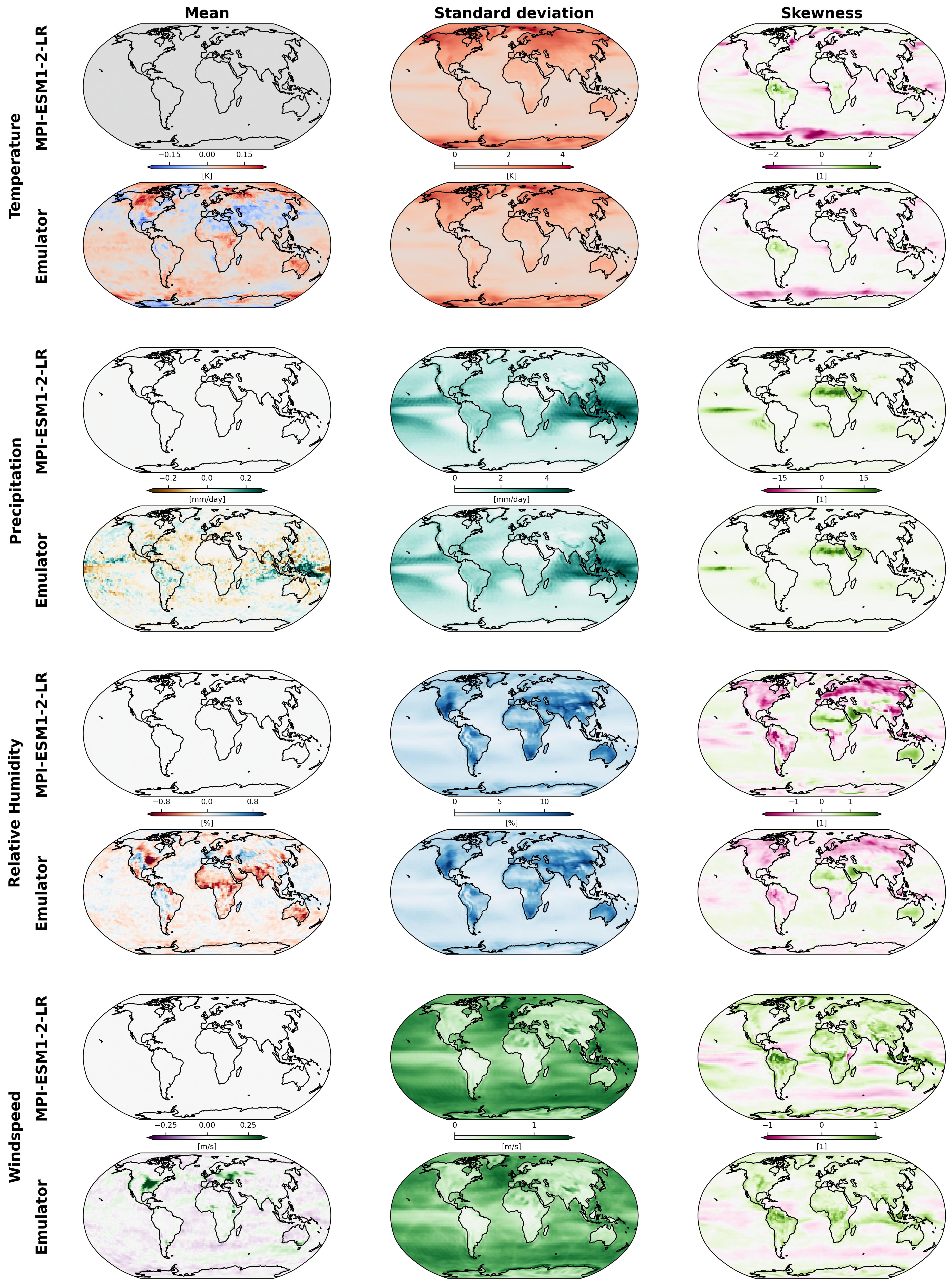}
    \caption{Comparison of moments from the MPI-ESM1-2-LR piControl run and the emulated distribution without forcing. Moments are estimated using anomaly data from all calendar months. Corresponding results for MIROC6 and ACCESS-ESM1-5 are provided in Section E of the supplementary.}
    \label{fig:moments}
\end{figure}

\subsection{Grid-cell distributions of variability}

We compare moment estimates using monthly fields from all calendar months altogether. Figure~\ref{fig:moments} shows that for all the emulated variables, the first three moments (mean, variance, skewness) of their emulated variability distribution provide a good match to the moments of the original climate model piControl run. In particular, the standard deviations align remarkably well. The emulator introduces some bias in the mean for all emulated variables. However, it generally remains minimal in magnitude relative to the standard deviation; that is, in individual realizations, this bias in the mean is not detectable and buried under variability. Skewness tends to be underestimated in some regions, but its spatial pattern and overall magnitude remain consistent with the ESM.

\begin{figure}[h]
    \centering
    \includegraphics[width=\linewidth]{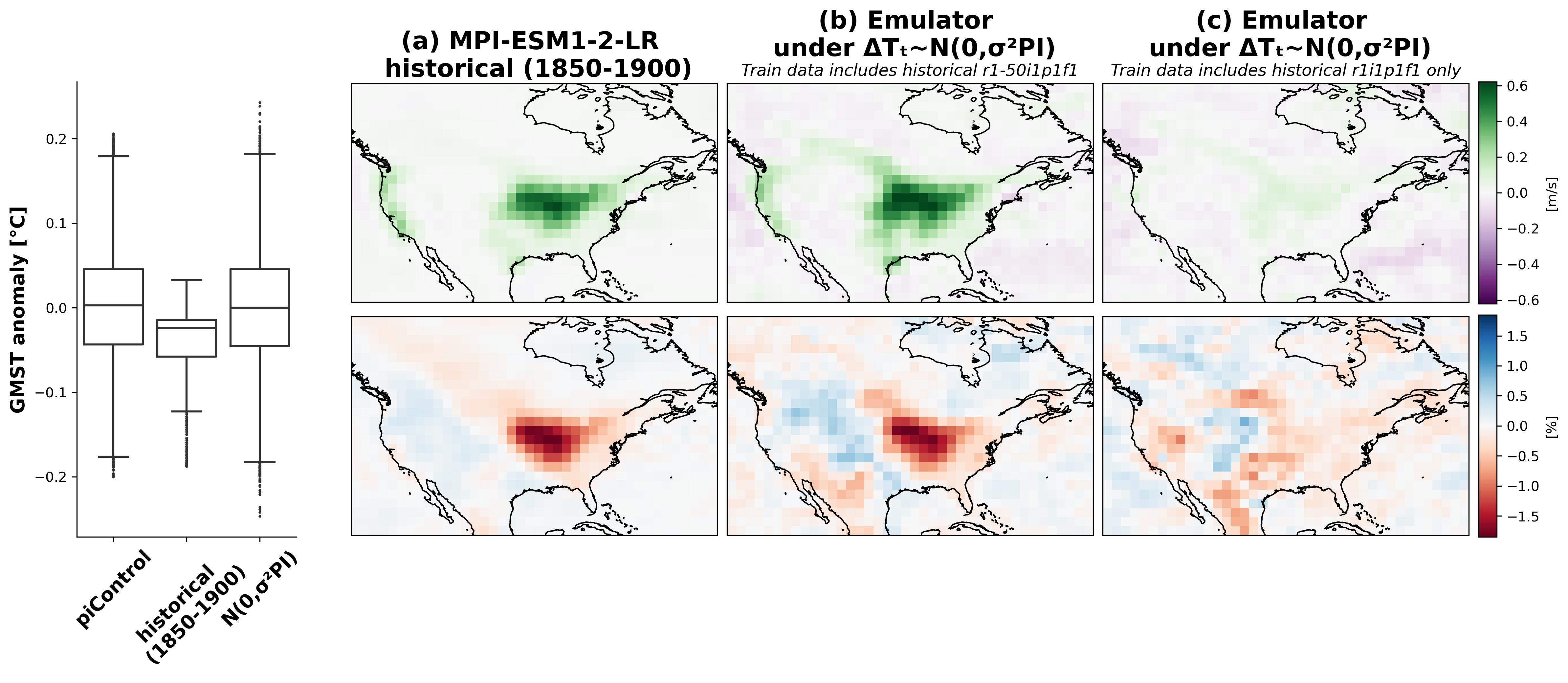}
    \caption{Left: Boxplots of GMST anomalies for the MPI-ESM1-2-LR piControl run, the MPI-ESM1-2-LR early historical period (1850–1900), and samples drawn from $\cN(0, \sigma_{PI}^2)$ used to generate the emulated ensemble. Right: Maps of mean monthly wind speed (top) and relative humidity (bottom) anomalies from (a) the MPI-ESM1-2-LR simulations of the early historical period, (b) the emulated ensemble when the training data includes all MPI-ESM1-2-LR members for the historial experiment, (c) the emulated ensemble when the training data includes a single member for the historial experiment.}
    \label{fig:overfitting}
\end{figure}

Some grid cells in the central US, as well as parts of Ukraine and western Russia, show biases in wind speed and relative humidity large compared to the magnitude of variability in these regions. Similar bias patterns appear in the emulations of the two other ESMs we considered. Figure~\ref{fig:overfitting} zooms in on this bias, focusing on the central US. It shows that the bias pattern (Pannel b) matches the regional anomalies in wind speed and relative humidity simulated by the ESM during the early historical period (1850–1900; Pannel a). This period accounts for roughly half of the training data at low GMST anomaly levels. Because GMST anomalies in the early historical and piControl runs overlap substantially, this suggests that the generative model is overfitting to spatial anomaly patterns from the historical simulations. Since ESM runs over the historical period are generally tuned to match observational records more closely, it is likely that this anomaly appears across many historical realizations and therefore becomes over-represented in the training data at these GMST levels. We find that withholding all but one ensemble member of the historical experiment from the training data eliminates this bias (Pannel c).

\begin{figure}[h]
    \centering
    \includegraphics[width=0.98\linewidth]{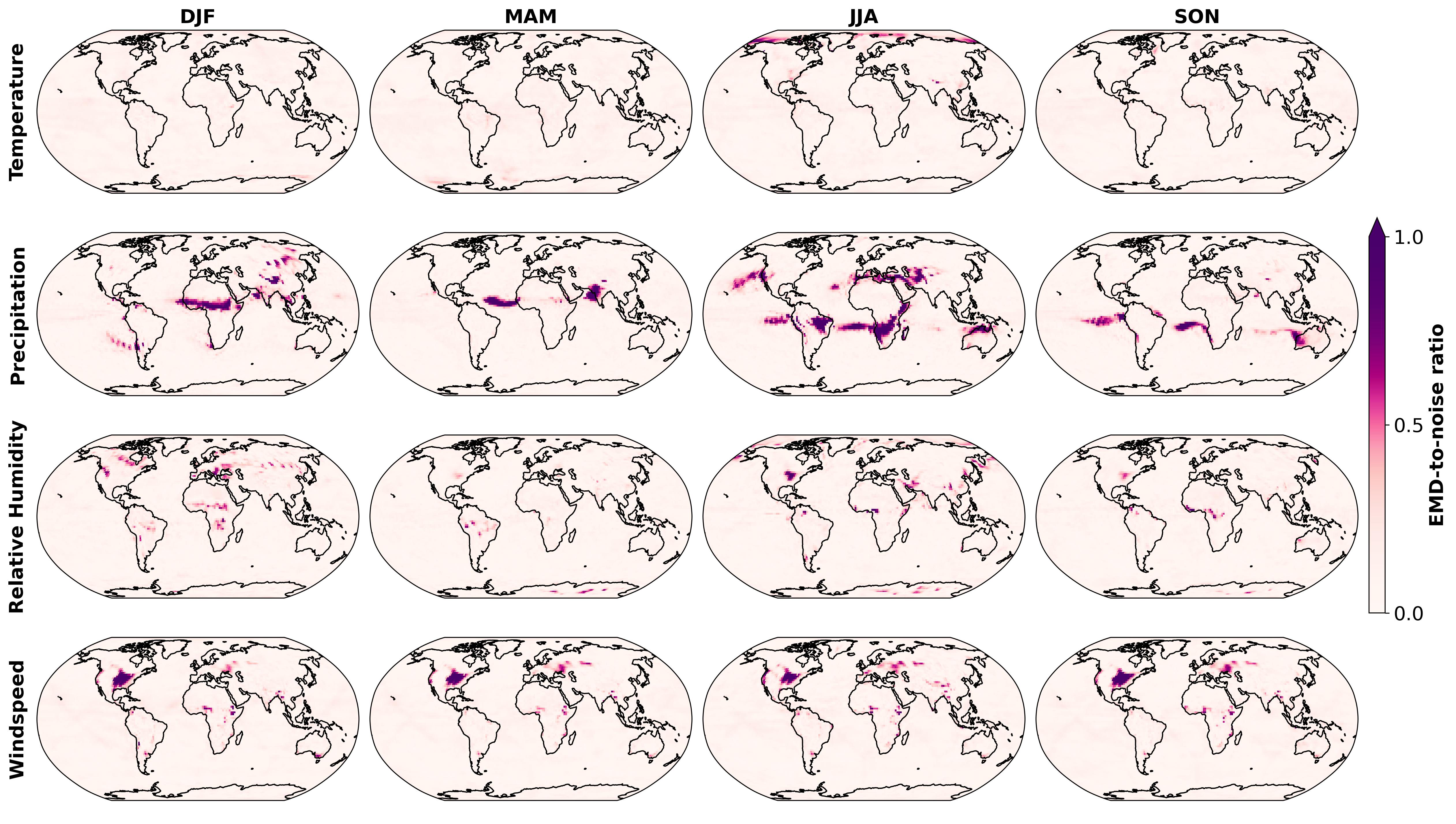}
    \caption{Spatial maps of grid-cell EMD-to-noise ratios between the MPI-ESM1-2-LR piControl run and the emulated large ensemble without external forcing. The reference standard deviation at each grid-cell $\sigma_{\texttt{gc,var}}$ is floored to $0.1$ to prevent spurious inflation of the ratio.}
    \label{fig:emd}
\end{figure}

Moving beyond individual moments of the emulated monthly variability, we now turn to the tougher test of comparing the entire emulated distribution to the climate model distribution. For each season, variable, and grid cell, we compute the EMD-to-noise ratio between the emulated distribution and the ESM piControl distribution and report it in Figure~\ref{fig:emd}. Overall, we find that shifts much smaller than the regional magnitude of variability are required to match the distributions at each grid cell. This means that, while discrepancies between the climate model output and the emulated distributions exist, they show in most regions a good agreement relative to the scale of internal variability in the climate model.

This is not true everywhere, and the maps identify regions where the approach fails, with an anomalously high EMD-to-noise ratio. The overfitting biases in wind speed and relative humidity in the central US and eastern Europe are clearly visible; as noted above, they can be mitigated by including a single member of the historical simulations in the training data. For the MIROC6 emulator (see supplementary Section F), several sharply delimited regions show high EMD-to-noise ratios in wind speed, which we attribute to a possible artifact in the model’s wind speed simulation.

For the three ESMs considered, we find that over a small number of grid cells, in particular over land, the emulator tends to generate \enquote{smoothed} distributions that exaggerate internal variability. Examples include the Arctic summer surface temperature, which corresponds to the overestimated monthly variability shown in Panel (c) of Figure~\ref{fig:emd-examples}, and spring precipitation over India shown in the supplementary Figure D.1. This phenomenon is most pronounced for the relative humidity in ACCESS-ESM1-5, where the emulator struggles with narrow distributions and smooths them excessively (see supplementary Figure F.4). While the black-box nature of neural networks makes it difficult to identify the cause of these discrepancies, literature on diffusion models shows they learn a smooth approximation to the score function that closely resemble the score of Gaussian mixture models~\citep{wang2024unreasonable, aithal2024understanding}. This provides a possible explanation for why distributions are smoothed out. Empirically, we find that increasing the number of parameters of the neural network used to estimate the score function largely mitigates this problem.

\begin{figure}[H]
    \centering
    \includegraphics[width=\linewidth]{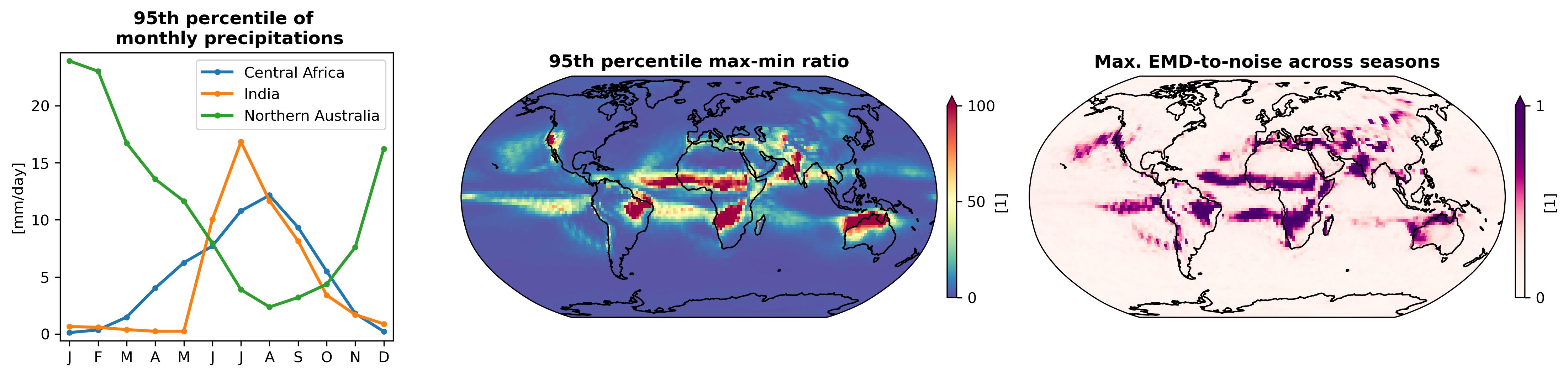}
    \caption{(Left) 95\textsuperscript{th} percentile of monthly precipitation in the MPI-ESM1-2-LR piControl run for three regions where the emulator fails to match the ESM distribution; (Center) Spatial map of the max–min ratio of the monthly 95\textsuperscript{th} percentile at each grid cell, highlighting regions with contrasts between months in dry or precipitating regimes; (Right) Max EMD-to-noise ratio across seasons.}
    \label{fig:precip-discrepancy}
\end{figure}

A more persistent challenge arises for precipitation across ESMs, where large EMD-to-noise ratios appear across seasons despite attempts at refining the diffusion model. We find that these discrepancies consistently occur in regions marked by some months with very low rainfall followed by months receiving susbtantial precipitation, e.g., regions that experience seasonal migration of the Intertropical Convergence Zone. Figure~\ref{fig:precip-discrepancy} shows the 95\textsuperscript{th} percentile of monthly precipitation in the MPI-ESM1-2-LR piControl, taken as an indicator of whether a given month belongs to a dry or precipitating regime. Taking the max–min ratio of this indicator across months identifies regions with pronounced seasonal contrasts, which align closely with those showing large EMD-to-noise ratios. This suggests that the diffusion model struggles to capture important seasonal distributional shifts in precipitation. The overestimation of variability in Arctic summer temperatures discussed earlier may reflect the same difficulty in representing important seasonal changes. A possible reason is it can be challenging for diffusion models to generate multimodal distributions from a unimodal Gaussian~\citep{xu2024disco}. This may be even harder if the modes are not equally represented~\citep{sehwag2022generating,qin2023class}. Pattern scaling alone may also not provide sufficient seasonal control. However we found empirically that adding additional input channels to the model for seasonal embeddings showed limited improvement.

\subsection{Cross-correlation across variables and grid-cells}

An important argument in favor of deep generative models is their promise to generate samples of multiple variables at every grid cell jointly. Unlike having separate emulators for each variable, such a joint model should preserve cross-correlations across variables and space. The emulator implicitly learns to represent correlations between all grid cells and variables simultaneously. For a climate field with 2$^\circ$ horizontal resolution and four variables, this amounts to more than 2.5 billion correlation terms. Examining every individual correlation is overwhelming; therefore, we focus on two diagnostics: cross-variable correlations at the grid-cell level, which speak to compound consistency, and spatial correlations within each variable, which highlight the scales of variability captured.

\begin{figure}[h]
    \centering
    \includegraphics[width=\linewidth]{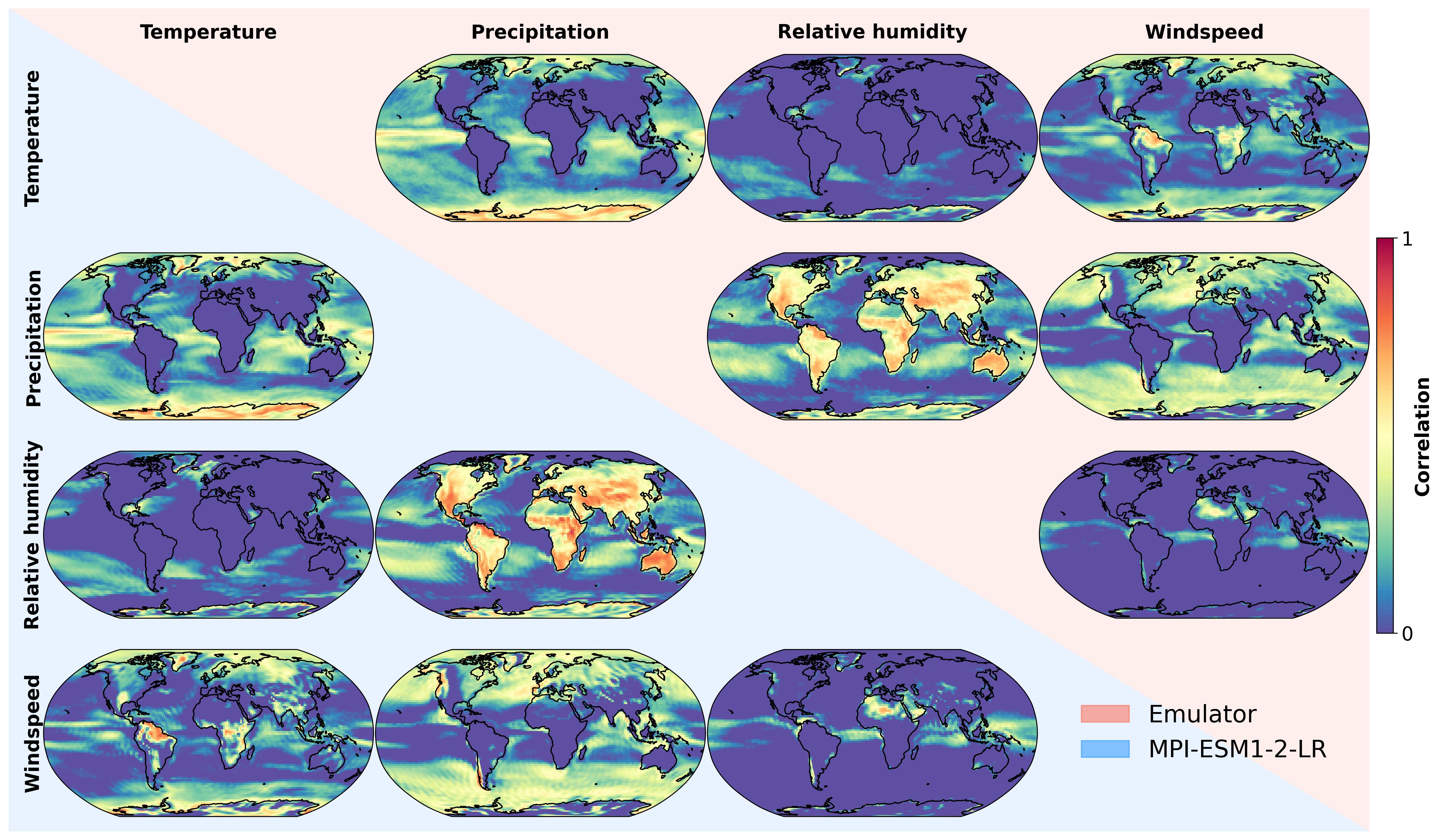}
    \caption{Spatial maps of grid-cell cross-correlations for all months between each pair of the four variables, shown for the MPI-ESM1-2-LR piControl run (bottom left) and for the emulated large ensemble (top right). Corresponding results for MIROC6 and ACCESS-ESM1-5 are provided in the supplementary Section G.} 
    \label{fig:crosscorrelations}
\end{figure}

We first examine the cross-variable correlations at individual grid-cells. An effective emulator should generate realizations that satisfy statistical and physical consistency across variables, and therefore open the door to studying projected compound events under alternative emission pathways. Figure~\ref{fig:crosscorrelations} compares the grid-cell cross-correlation between the variables obtained from the proposed generative emulator, with those estimated from the MPI-ESM1-2-LR piControl run. It shows the variables' cross-correlations from emulator matches those from the reference ESM it aims to emulate. Because the current emulator generates monthly snapshots and does not incorporate temporal dependencies, its utility for temporal compounding event studies is limited (daily time series are typically required). However, these results suggest that generative models can reproduce statistical consistency across variables, and therefore have potential in this area. With additional compute to condition the emulator on past states and generate daily samples, the emulator can be extended to fit this purpose, following for example the work of \citet{bassetti2024diffesm}.

\begin{figure}[H]
    \centering
    \includegraphics[width=\linewidth]{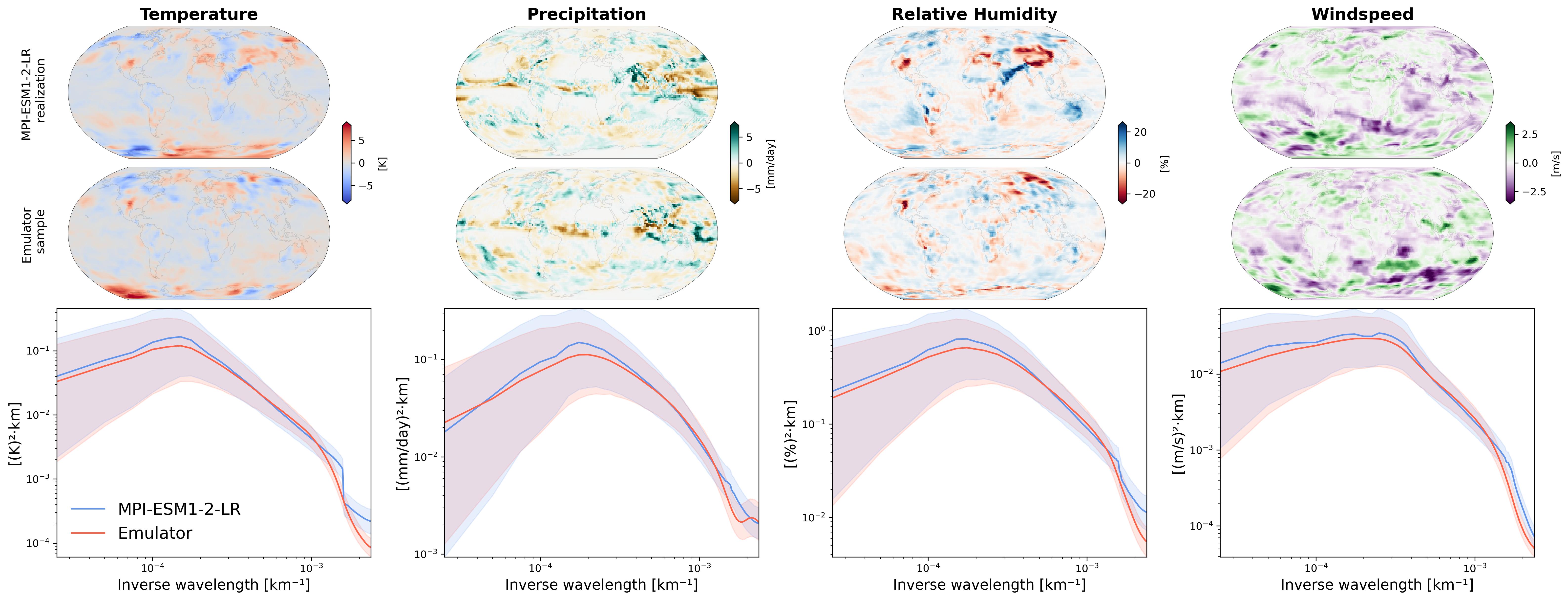}
    \caption{Top: Examples of individual realizations from the MPI-ESM2-1-LR and emulator for the same randomly selected month (July) illustrating the spatial structure of realizations. Bottom: global zonally averaged spherical power spectra from an ensemble of individual realizations for each variable. Spectral densities are computed for all available data in the MPI-ESM2-1-LR piControl run and the emulated large ensemble without forcing. Full lines indicate the mean power density and shaded area the 95\% interquantile range across realizations.}
    \label{fig:raspd}
\end{figure}

We next examine spatial correlations within each variable by studying the spatial structure of individual emulator realizations. For each variable, we compute the global, zonally averaged spherical power spectra of individual realizations and form an ensemble of power spectral densities. Figure~\ref{fig:raspd} shows an example of a spatial sample for each variable along with the ensemble power spectral density. The emulator's mean power spectrum is slightly misaligned with that of the ESM, and this difference is statistically significant. For large spatial scales, however, variability in power density across realizations is sufficiently large that these differences are unlikely to be perceptible in individual samples. This is no longer the case as we get closer to the ESM native grid resolution. It indicates that the emulator should be able to reproduce synoptic-scale structures such as heat domes or monsoons, but is less reliable at replicating ESM output features confined to only a few grid cells.

At scales close to the numerical meshgrid, ESM projections themselves are influenced by parameterization biases, and impact models typically do not treat them as perfectly accurate. They rely on bias correction, downscaling, and multi-model ensembles to correct inputs or explicitly incorporate uncertainty, thereby ensuring that sensitivities to climate model errors do not mislead conclusions~\citep{falloon2014ensembles, maraun2016bias, lange2019trend}. In this context, an emulator that exhibits fine-scale inaccuracies that remain within the tolerated envelope of input uncertainty should be just as useful as the original ESM for impact studies.

\section{Emulation of forced simulations}\label{section:forced}

We now evaluate if the emulator can reproduce the climate model output distribution in a climate change scenario forced with anthropogenic greenhouse gas emissions. The goal is to assess how well it performs when $\Delta T_t \neq 0$. An emulator is again trained for each ESM on the piControl, historical, SSP1-2.6, and SSP5-8.5 runs

Because emulators are intended to explore alternative emission pathways not run by climate models, we evaluate them on scenarios that were excluded from the training set for this purpose. We choose SSP2-4.5, an intermediate scenario with emission peaking by mid-century, and SSP3-7.0, a high-emission scenario with fragmented climate policies because of regional rivalry. These two scenarios lie between the strong mitigation and very high forcing trajectories used for training, and therefore test the emulator’s skill at projecting within the range of ScenarioMIP forcings. The emulator is an interpolation algorithm and is not expected to have skill in extrapolating beyond the training range.

For each prediction scenario, we generate a large ensemble of emulated climates conditioned on the GMST time series. At a given time $t = (\text{year, month})$, the large ensemble is generated by computing the GMST anomaly $\Delta T_t$ from the reference climate model output, and then sampling from $q(x \mid \Delta T_t)$ 50 times.

\subsection{Forced trends}

We first evaluate whether the emulator is able to reproduce the forced climate change trends simulated by the ESM. To assess this, we compute mean annual anomalies for each variable over the SSP3-7.0 scenario using ESM output data and an emulated large ensemble. We choose SSP3-7.0 because it has a stronger forcing, so the climate change signal is expected to emerge more clearly.

\paragraph*{Temperature and relative humidity trends} We study the emulated trends averaged for four characteristic regions with different warming rates following \citet{giani2025origin}: land, tropical ocean (ocean between 10$^\circ$S-10$^\circ$N), Southern Ocean (ocean poleward of 55$^\circ$S), and Arctic region (poleward of 80$^\circ$N). Land warms faster than the oceans due to a lower heat capacity and limited evaporative cooling, which also results in a drop in relative humidity~\citep{byrne2016understanding, byrne2018trends}. The Arctic has the fastest warming rate because of positive feedbacks, and displays high variability~\citep{previdi2021arctic}. The greater heat capacity and unlimited moisture supply in the oceans leads to a slower warming and smaller changes in relative humidity per degree Kelvin~\citep{byrne2018trends}. The Southern Ocean warming is further muted due to strong wind-driven upwelling of cold waters and melting of ice shelf~\citep{armour2016southern, dong2022antarctic}.

\begin{figure}[htbp]
    \centering
    \includegraphics[width=\linewidth]{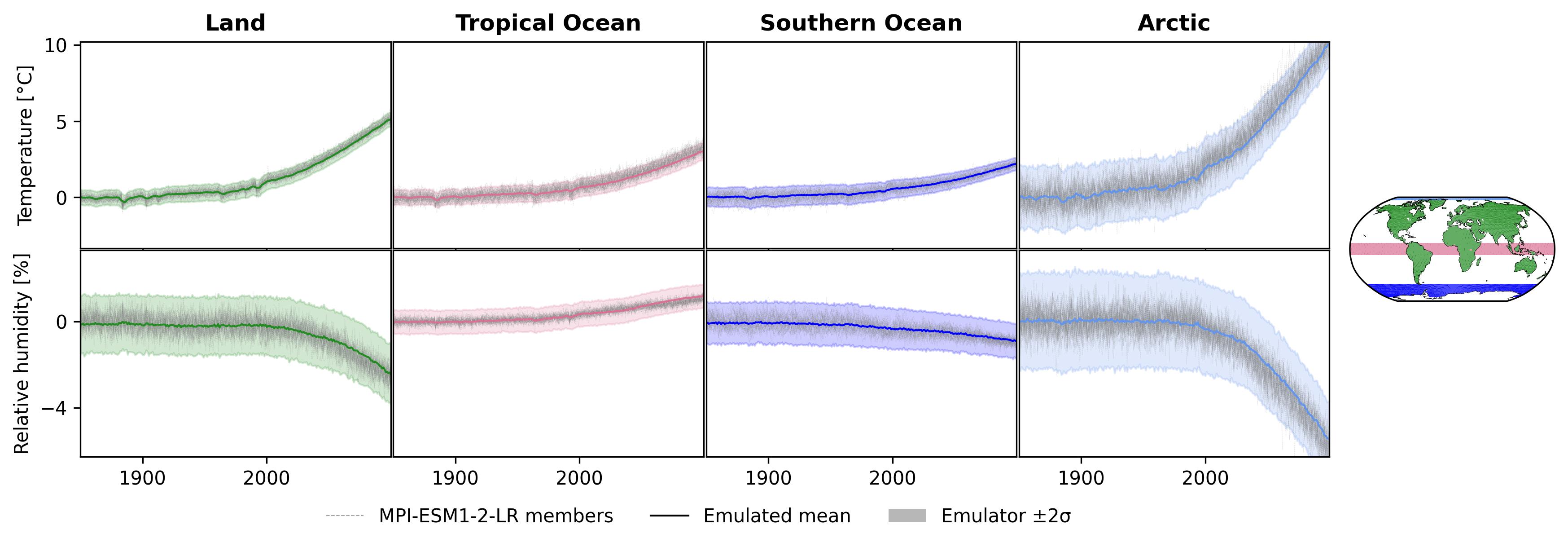}
    \caption{1850-2100 evolution of the emulated mean annual anomaly for temperature (top) and relative humidity (bottom) under SSP3-7.0 in different world regions, shown in the world map to the right. Full lines represent the emulated mean, overlayed on top of the MPI-ESM2-1-LR ensemble members (gray). Spread represents 2 standard deviations from the emulator mean.}
    \label{fig:tas-hurs-trends}
\end{figure}

In Figure~\ref{fig:tas-hurs-trends}, the mean emulated trend is shown together with the range defined by two standard deviations above and below the mean. They are overlayed on the ESM ensemble member trajectories for the same region. We find that the emulator reproduces the trends associated with the simulated warming fingerprints of each region, in agreement with the ESM trajectories. The emulator can adapt to differences in magnitude and sign of regional fingerprints depending on the ESM on which it is trained. For example, it successfully reproduces the opposite trends in Southern Ocean relative humidity anomalies between the MPI-ESM1-2-LR and MIROC6 projections (shown in the supplementary Section I).

The figure also reveals a modest underestimation of mean temperature and relative humidity anomalies by the end of the century in some regions. This effect is most pronounced in the Arctic for ACCESS-ESM1-5 and MIROC6, where the emulator does not fully capture the late-century anomaly range. For relative humidity, the emulator further tends to overestimate variability. Nonetheless, the ESM ensemble members are generally well captured within two standard deviations of the emulated large ensemble. This shows the emulator has skill in reproducing the statistics of forced trends for a scenario absent from its training data.

\begin{figure}[htbp]
    \centering
    \includegraphics[width=\linewidth]{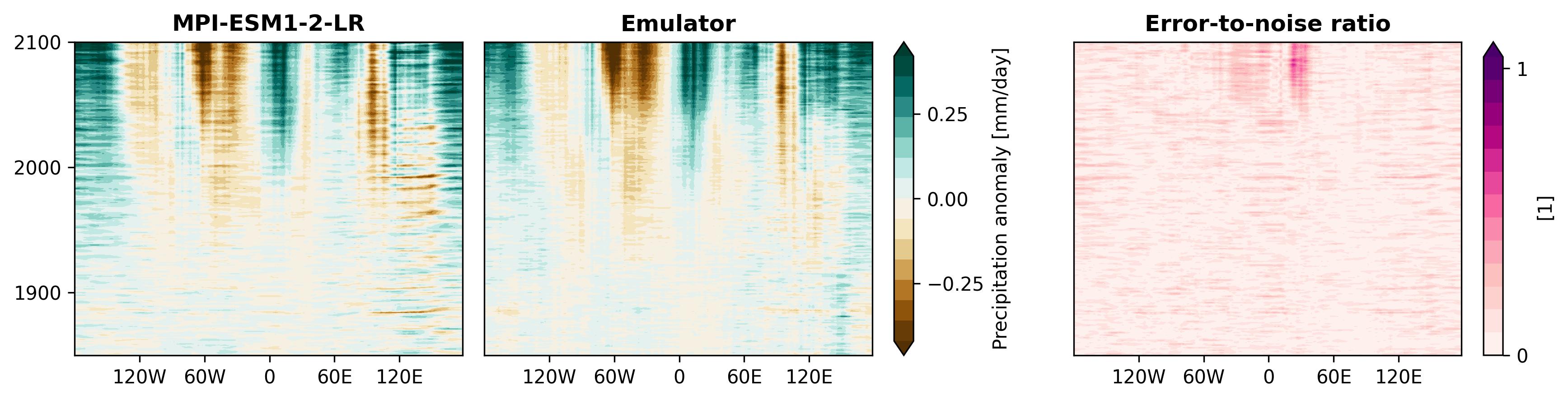}
    \caption{Hovmöller diagrams of 30$^\circ$N–30$^\circ$S annual mean precipitation anomalies under SSP3-7.0 for MPI-ESM1-2-LR (left) and the emulated large ensemble (center). The right panel shows the error normalized by the MPI-ESM1-2-LR standard deviation across ensemble members.}
    \label{fig:hovmoller}
\end{figure}

\paragraph*{Precipitation trends} For precipitation, we focus on the tropical belt (30$^\circ$N–30$^\circ$S), where most of the variations in global mean precipitation arise, in association with the circulation changes that dominate future hydro-climate projections~\citep{trenberth2011changes}. Figure~\ref{fig:hovmoller} shows a longitude-time view of the annual mean tropical precipitation anomaly under SSP3-7.0, comparing the ESM ensemble mean with the emulated ensemble mean.

The emulator reproduces the large-scale anomaly patterns seen in the ESM, with wetter conditions over the tropical Pacific (120$^\circ$E–120$^\circ$W) and drier conditions over the Amazon rainforest ($\sim$60$^\circ$W)~\citep{lee2021future}. This reflects the \enquote{wet-get-wetter/dry-get-drier} feedback~\citep{held2006robust} in which regions already prone to heavy rainfall experience more precipitation, while evaporating areas such as the Amazon basin tend to dry. Note that part of this drying signal may reflect a known bias of the MPI-ESM1-2-LR model, which underestimates rainfall over tropical land~\citep{mauritsen2020tuning}. Some discrepancies in magnitude and smaller-scale structure of the emulated and ESM projected precipitation anomaly persist. To gauge their importance, we compute the absolute difference between the two anomaly fields and normalize it by the noise magnitude. The noise is taken as the ESM internal variability estimated across ensemble members. The resulting error-to-noise map in Figure~\ref{fig:hovmoller} shows ratios close to zero across times and longitudes. This suggests that the mismatch in the emulated mean forced trend is small compared to natural variability, and would be hard to detect in any single realization.

\begin{figure}[htbp] 
    \centering
    \includegraphics[width=\linewidth]{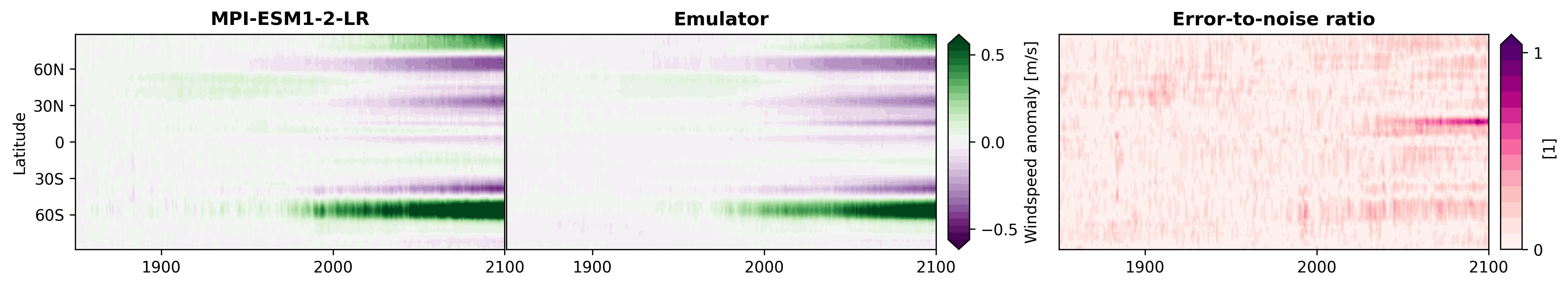}
    \caption{1850-2100 evolution of the annual zonal mean wind speed anomalies under SSP3-7.0 for MPI-ESM1-2-LR (left) and the emulated large ensemble (center). The right panel shows the error normalized by the MPI-ESM1-2-LR standard deviation across ensemble members.}
    \label{fig:westerlies}
\end{figure}

\paragraph*{Wind speed trends} We compute the evolution of the zonally averaged surface wind speed anomaly from the ESM output and emulated large ensemble. Zonal averages emphasize shifts in the westerlies and trade winds. The Southern Hemisphere westerlies are expected to shift poleward, in association with a weakening of low-level winds over the southern parts of South America, southern Africa, and Australia~\citep{lee2021future}. Arctic wind speed is projected to strongly increase as surface roughness decreases with ice melting and the boundary layer becomes warmer and less stratified~\citep{mioduszewski2018diminishing}. On the other hand, the response in Antarctica is expected to be less robust, as near-surface wind changes are more complex and regionally variable~\citep{davrinche2025future}. Figure~\ref{fig:westerlies} shows that, like for precipitation, the emulated wind speed anomaly trends match well the trends computed from the ESM outputs, and are largely able to reproduce the aforementioned trends. Notably, the emulator reproduces shifts in wind patterns despite being conditioned on pattern scaling, which assumes a fixed pattern of change for surface temperatures.

\subsection{Time of Emergence}

The time of emergence (ToE) refers to the moment when a forced climate change signal becomes detectable and rises above the background variability~\citep{hawkins2012time}. Traditionally, this has been quantified using the signal-to-noise ratio, where the signal is the long-term trend from a single model's projection and the noise is internal variability estimated from pre-industrial control runs or large ensembles of the same model~\citep{barnett1987detecting, giorgi2009time}. This is an important concept for impact assessment as it signals when the Earth's climate becomes unequivocally different from what was previously experienced. A recurrent assumption in the aforementioned studies of emergence is that the variance of internal variability is approximately stationary.

To verify whether the emulator can reproduce the ToE simulated by the ESM, we select SSP3-7.0 as a scenario with a sufficiently strong greenhouse gas forcing so that we can expect the projection simulated by the ESM to exhibit regional emergence for all four variables considered. We emulate a large ensemble under the SSP3-7.0 GMST trajectory and compute for each variable the first year from which the signal-to-noise ratio of yearly averaged variables becomes greater than 2 for three years in a row. The noise magnitude is estimated as the anomaly standard deviation under the ESM piControl run.

\begin{figure}[htbp]
    \centering
    \includegraphics[width=\linewidth]{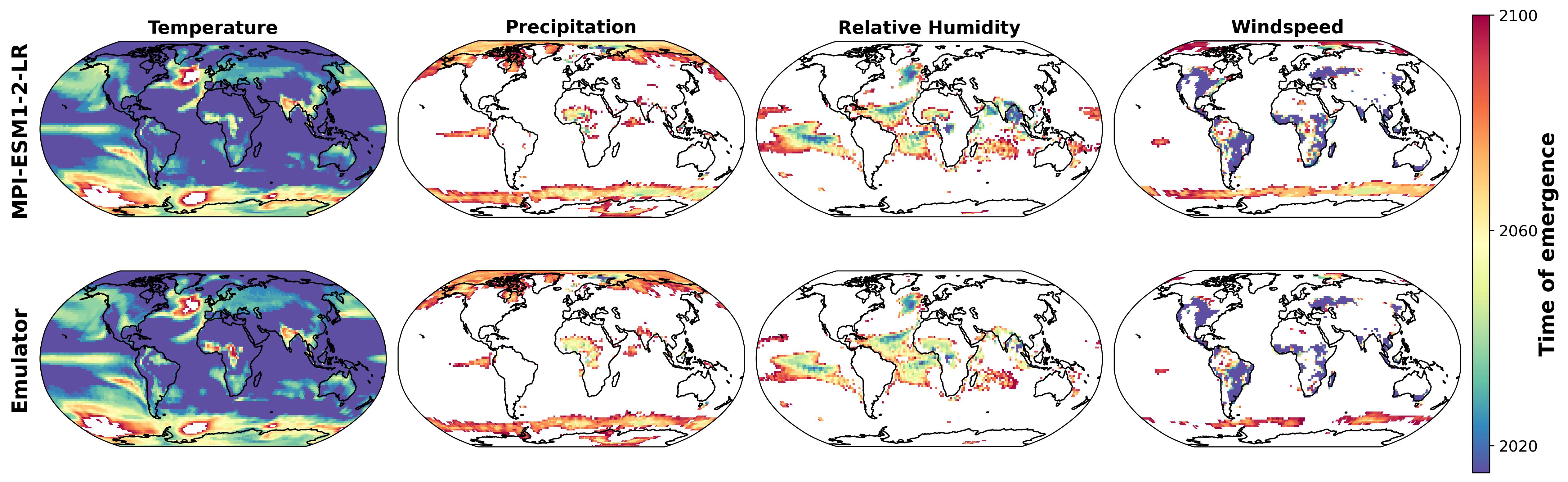}
    \caption{Time of emergence of the annually averaged anomalies computed for the MPI-ESM1-2-LR simulated ensemble and for the emulated ensemble under SSP3-7.0}
    \label{fig:ToE}
\end{figure}

Figure~\ref{fig:ToE} shows that the emulated ensemble is largely capable of recovering a pattern of emergence that matches the ESM, with later emergence of the temperature signal in the Antarctic due to higher variability, emergence of precipitation at high latitudes and in some tropical regions~\citep{ranasinghe2021climate}, or emergence from the poleward shift in southern westerlies. We include in Section D of the supplementary results for SSP2-4.5, which shows that as we emulate a lower forcing scenario, emergence for all variables either occurs at a later year or never occurs, in accordance with the ESM response. While our definition of ToE is somewhat arbitrary, and more rigorous estimates of emergence have been proposed~\citep{li2017quantifying, rivoire2024observational}, this diagnostic already reflects how good the emulator is at reproducing the signal-to-noise ratio from the climate model output.

\subsection{Distributional change with warming}

Warming is anticipated to affect not only the mean response, but also higher-order characteristics of the anomaly distributions of climate variables. We examine how accurately the emulator can reproduce these distributional changes with warming. We focus on SSP2-4.5 because it is a common pathway in impact assessment, and allows to evaluate emulator skill beyond the high-forcing conditions of SSP3-7.0.

\begin{figure}[H]
    \centering
    \includegraphics[width=\linewidth]{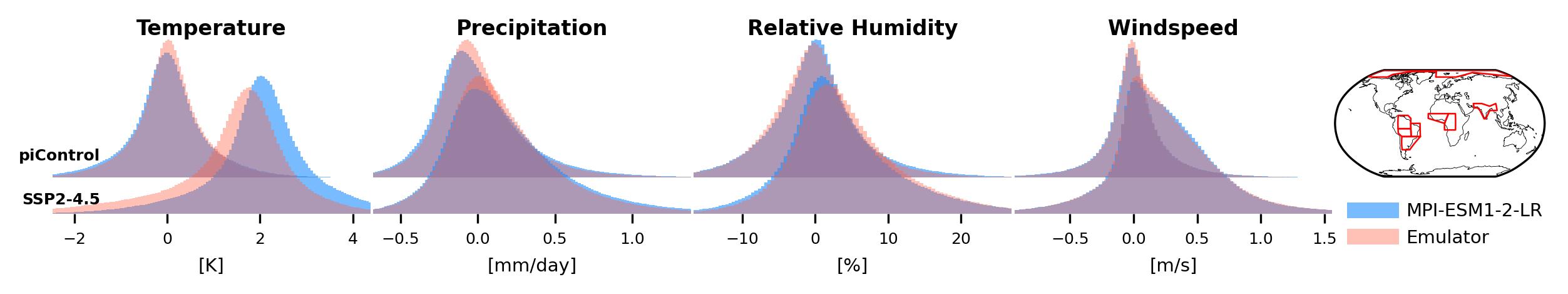}
    \caption{Distributional change of the climate model output and emulated large ensemble between pre-industrial era and end of century (2080-2100) under SSP2-4.5. Distributions are binned using monthly data over a selection of AR6 regions outlined in red on the right: central and western Africa for temperature, Arctic Ocean for precipitation, south Asia for relative humidity, and South America for wind speed.}
    \label{fig:joy}
\end{figure}

Figure~\ref{fig:joy} compares anomaly distributions for the four emulated variables in two states: the pre-industrial control climate and the end of the century under SSP2-4.5. The distributions are plotted for a selection of regions where the climate change signal has emerged in the ESM outputs under SSP2-4.5. The emulator seems to reproduce the main distributional shifts seen in the ESM output. While some bias is visible for emulated temperature anomalies over tropical Africa, the bulk of each distribution appears well captured by the end of the century for the regions considered. In particular, the emulation of the change in skewness for wind speed over South America and relative humidity over India is appreciable, highlighting its ability to reproduce higher-order distributional shifts that Gaussian distributions cannot represent.

\begin{figure}[H]
    \centering
    \includegraphics[width=\linewidth]{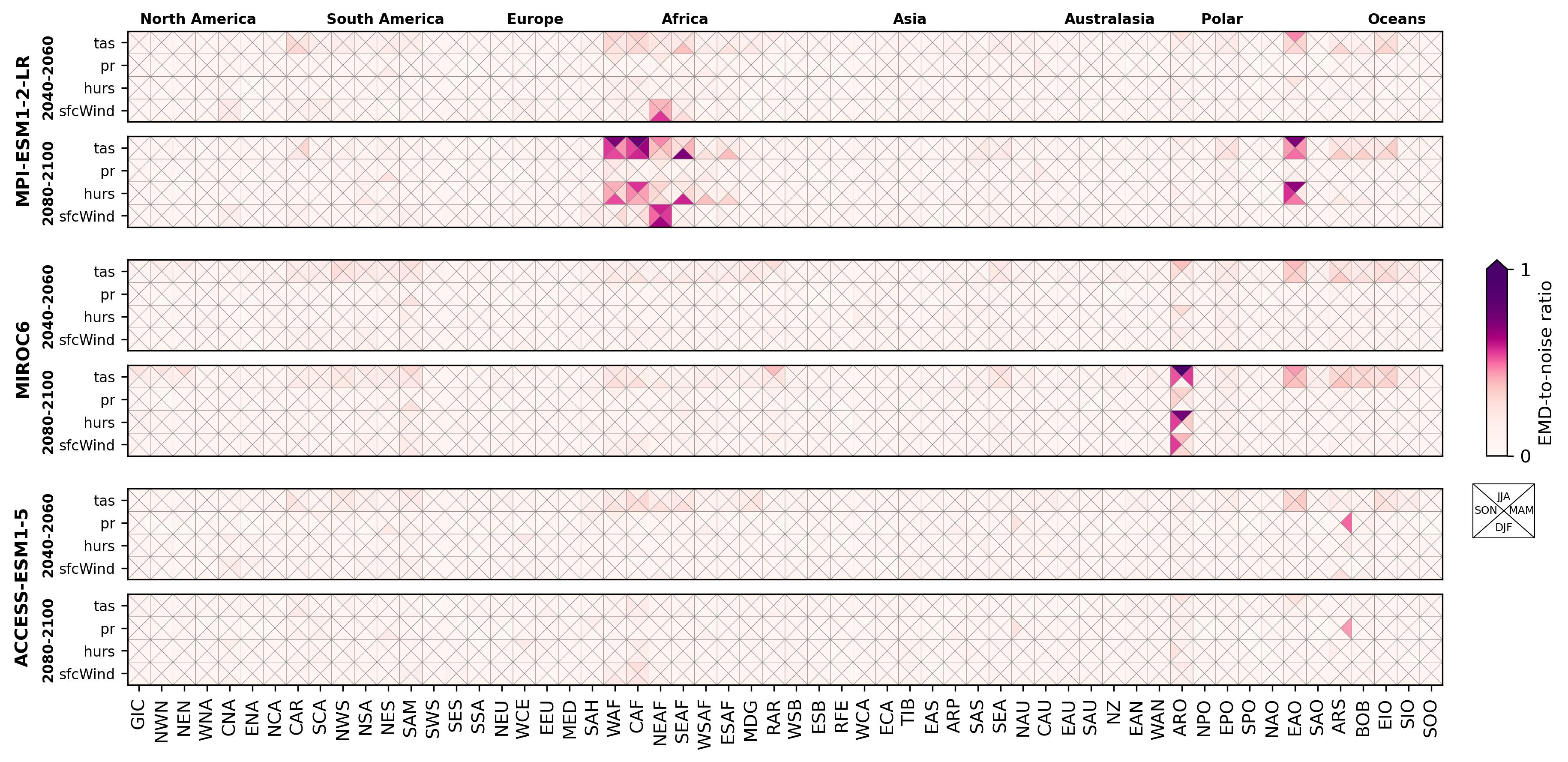}
    \caption{EMD-to-noise ratio under SSP2-4.5 for each ESM, variable, season and AR6 region. Results are shown for two periods of the scenario: mid-century (2040–2060) and end-century (2080–2100). The reference standard deviation is floored to 0.1 to prevent spurious inflation of the ratio.}
    \label{fig:portait}
\end{figure}

\begin{figure}[htbp]
    \centering
    \includegraphics[width=\linewidth]{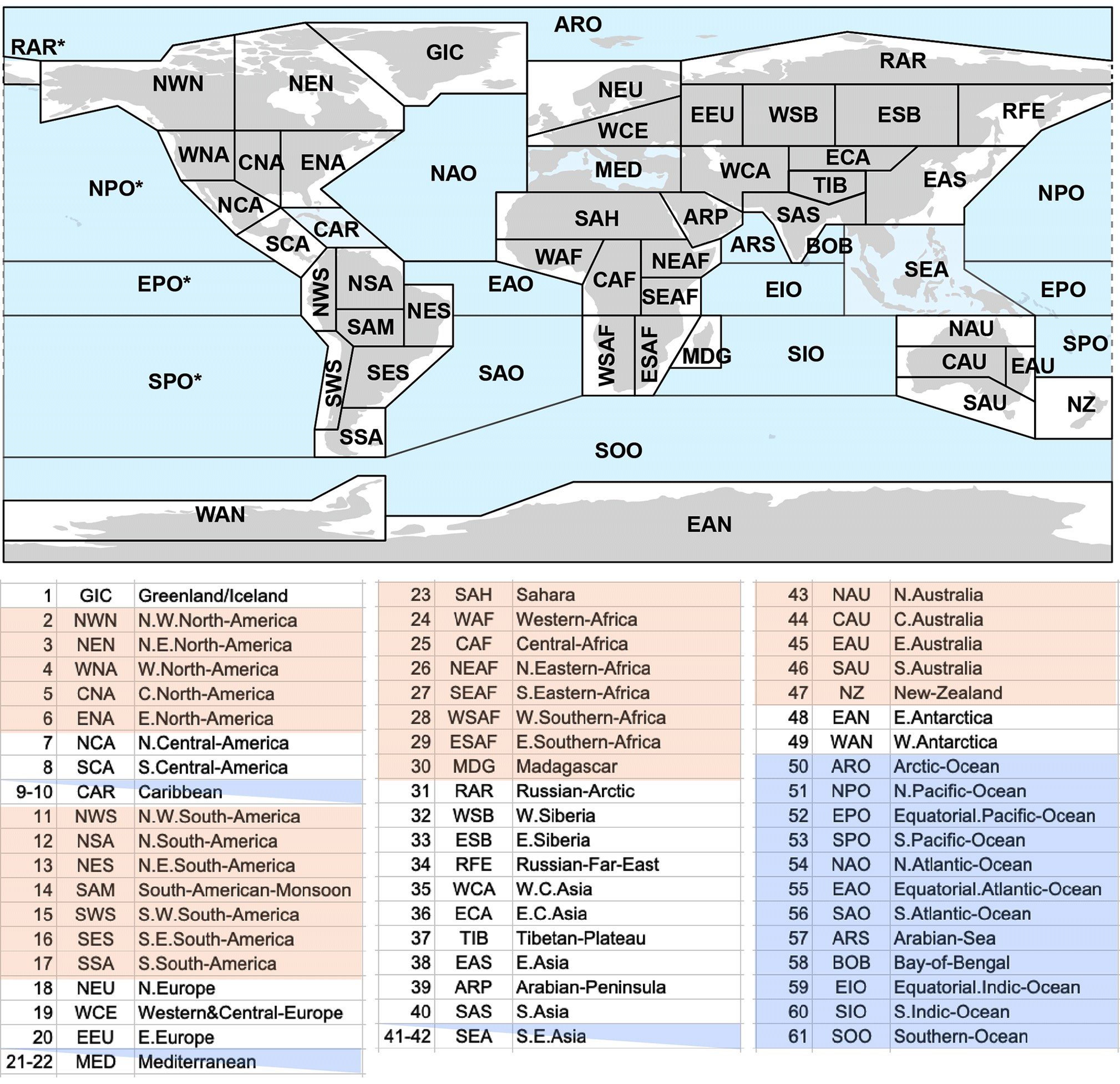}
    \caption{Map of AR6 regions with acronyms. Source: \citep[Figure 1]{iturbide2020update} licensed under CC BY 4.0.}
    \label{fig:ar6}
\end{figure}

In Figure~\ref{fig:joy}, the probability masses that need to be rearranged to bring ESM and emulator projections into agreement correspond to the areas shaded in blue and red. Visually, they appear small in comparison with the range covered by the variability. To provide a comprehensive assessment of this, we compute --- separately for each ESM, AR6 region, season, and for mid-century and end-century periods under SSP2-4.5 --- the EMD-to-noise ratio between the emulated and the ESM output distribution. A map of AR6 regions~\citep{iturbide2020update} with acronyms is provided in Figure~\ref{fig:ar6}. The results in Figure~\ref{fig:portait} support that in a majority of cases, the distributions align well relative to the scale of internal variability, with an EMD-to-noise ratio well below 0.5. We find that tropical Africa for the MPI-ESM1-2-LR model, including the bordering Atlantic ocean region, is in fact one of the few exceptions. It is also a region where errors in emulating this ESM have been reported in previous work~\citep{nath2022mesmer, geogdzhayev2025eof}. This is possibly linked to dynamic vegetation feedbacks in the MPI-ESM1-2-LR land model, which can display a nonlinear response on the near-surface climate in Africa that does not easily correlate with GMST~\citep{baudena2015forests, reick2021jsbach}. Additional mismatches in distribution are found in the Arctic Ocean for MIROC6, and the highest discrepancies generally occur late in the century, when SSP2-4.5 differs most from the training data.

\subsection{Extreme tails with warming}

Record-shattering events, whether observed or simulated, sit in the far tail of the climate distribution for any given year. They contribute to natural disasters that incur enormous human and economic losses. Although ESMs still have limitations in representing these extremes, they remain a useful indicator of future forced changes in extremes, while future advances in Earth system modeling will likely improve the representation of extremes in climate projection. Therefore, it is desirable for a climate model output emulator to accurately reproduce the most extreme events projected by ESMs under climate change.

\begin{figure}[htbp]
    \centering
    \includegraphics[width=\linewidth]{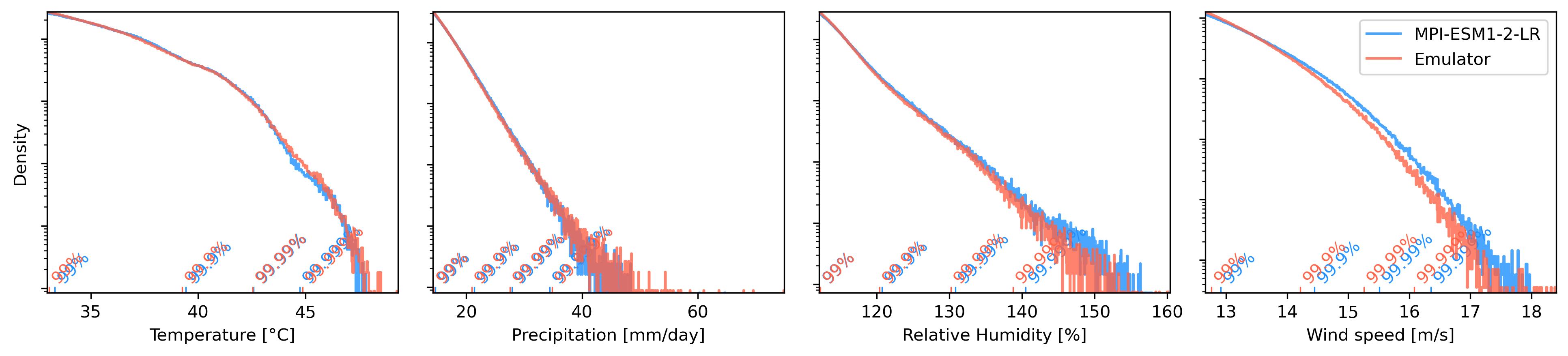}
    \caption{Extreme tails of the MPI-ESM1-2-LR output distribution and the distribution from the emulated large ensemble. Tails are binned from all grid cell values, ensemble members, and months over the 2080-2100 period under SSP2-4.5. Positions of the 99\%, 99.9\%, 99.99\%, and 99.999\% quantiles are reported on the x-axis for the ESM (blue) and emulator (red). Near-surface relative humidity values can spuriously exceed 100\% in ESM outputs~\citep{ruosteenoja2017surface}.}
    \label{fig:tails}
\end{figure}

Because extreme tail events carry little probability mass, the EMD is insensitive to their misalignment. We therefore prefer to evaluate biases in extreme tail quantiles at the end of the century. We convert back the anomalies into absolute values and compare the tails of the climate model output distribution and emulated ensemble distributions over 2080-2100 under SSP2-4.5. Distributions are binned by taking all grid cell values, ensemble members, and months, and we report in Figure~\ref{fig:tails} their tails beyond the 99\% quantile. The emulator shows good agreement with the ESM in the highest extremes for all variables, with modest underestimation of the most extreme quantiles for relative humidity and wind speed. The only notable bias we find is an underestimation of the extreme tails in relative humidity for ACCESS-ESM1-5. Working with monthly averages helps with emulating the tails of the distributions: time-averaging pulls extremes closer to the bulk of the distribution, making them easier for the emulator to match. Monthly extremes have relevance for seasonal risk assessments, for example, in simple crop-yield models or monthly water-balance models~\citep{ray2015climate, bock2017us}, and these results show the proposed emulator has skill to support them. Extreme tails are likely to become more challenging to emulate at daily or hourly resolution, which are the scales at which impact assessments matter most, and we leave this challenge for future work.

\begin{figure}[htbp]
    \centering
    \includegraphics[width=\linewidth]{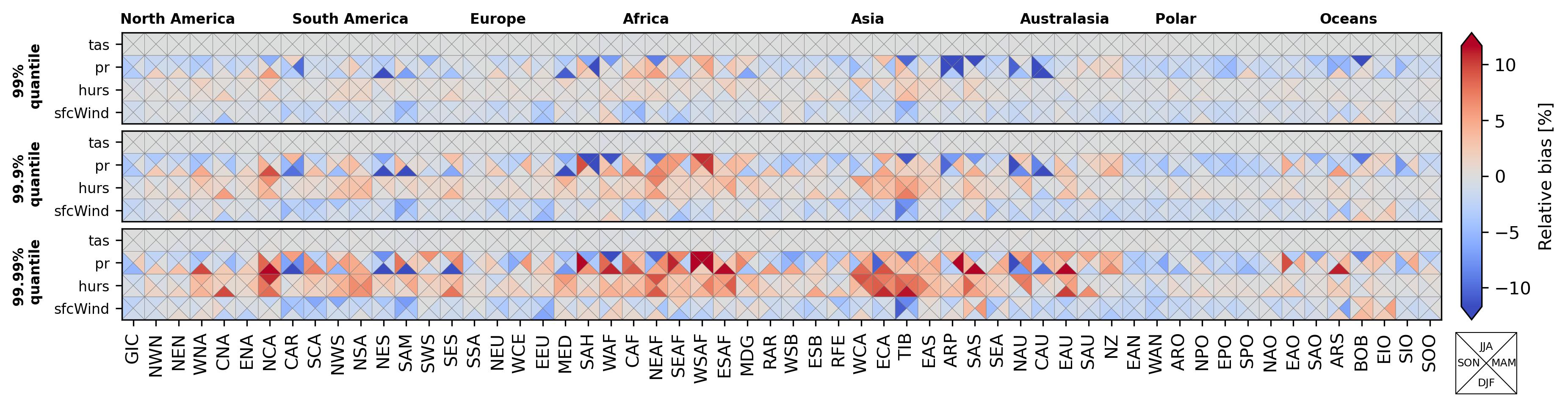}
    \caption{Relative bias in the extreme tail quantile between the MPI-ESM1-2-LR output and the emulated large ensemble for each AR6 region and season under SSP2-4.5. Values are reported as relative deviations with respect to the ESM quantile estimate, where red/blue is an overestimation/underestimation by the emulator.}
    \label{fig:quantile-bias}
\end{figure}

The histogram in Figure~\ref{fig:tails} bins values from all grid cells and months indiscriminately. This raises the question of whether the model gets the most extreme values right, at the right location, and in the right month. To assess this, we compute the relative deviation in tail quantiles for each AR6 region and each season under SSP2-4.5 and report the results in Figure~\ref{fig:quantile-bias}. We find that the bias in extreme tail quantiles from the emulator is mostly bounded within 10\% of the ESM corresponding quantile value. For temperature, relative humidity, and wind speed, the emulator matches the ESM’s tails in most regions and seasons, with occasional deviations as we move to more extreme quantile levels. The bias is larger at higher quantiles since rarer events provide less training signal. This is most visible for relative humidity in ACCESS-ESM1-5, where the relative bias increases most at the upper tails. Precipitation proves to be more challenging: it displays the largest underestimation of bias in the 99\% quantile for multiple regions, and upper tails show pronounced overestimation for a large number of regions and seasons. We test the statistical significance of biases in tail quantile estimates relative to their variability and find that they are significant in a majority of cases, with detailed results included in the supplementary Figure D.3. This suggests that even if errors in the extreme tails are small in magnitude, they still constitute a systematic error introduced by the emulator in its representation of tails.

\section{Discussion}\label{section:discussion}

\subsection{On the utility of emulating biased climate models}

One critique of climate model emulation is that many ESMs exhibit known biases, particularly in their representation of tails and fine-scale variability --- hence the large research effort devoted to model tuning and bias correction, and the constant drive to increase models' resolution. Therefore, if parts of the simulated distribution are known to be unreliable, it is reasonable to question whether emulating them is even useful. We argue that this critique is misplaced for two reasons.

First, while imperfect, ESMs have undeniably allowed us to advance our understanding of the climate system and predict its response to greenhouse gas emissions, as demonstrated in retrospective studies~\citep[e.g][]{hausfather2020evaluating}. Where biases exist, many applications already treat ESM outputs with caution. For example, bias correction, downscaling, and multi-model ensembles are standard tools in impact assessment workflows to adjust inputs and explicitly represent uncertainty, so that conclusions are not misled by ESM biases~\citep{falloon2014ensembles, maraun2016bias, lange2019trend}. In practice, this means emulators can be used directly where ESMs are reliable, and postprocessed through the same existing pipelines where they are not.

Second, the development of climate model emulators is also largely a methodological endeavor that is ESM-agnostic. As such, when better climate models are available, the same frameworks can be used to emulate them with minimal adjustment. In fact, we can already envision adding gridded reanalysis products in the training of emulators so that they match observed climate distributions better than ESM outputs~\citep{brenowitz2025climate}. For methodological development, as in this work, evaluating an emulator’s ability to capture future changes requires future projections, which justifies training and validation on CMIP6 archives. Once trained on CMIP6 simulations, however, the emulator can be fine-tuned on reanalysis products or available observations using transfer learning or nudging approaches to reduce model bias before practical use~\citep{immorlano2025transferring, wang2025gen2}.

\subsection{On the GMST forcing assumption}

\begin{figure}[htbp]
    \centering
    \includegraphics[width=\linewidth]{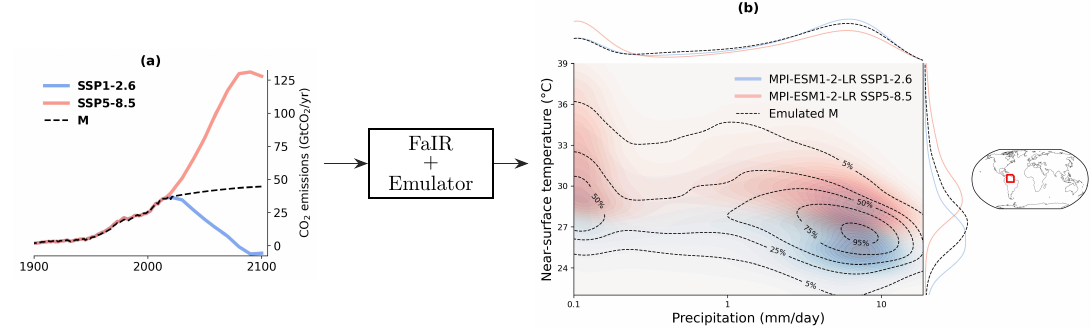}
    \caption{(a) Annual CO\textsubscript{2} emissions under SSP1-2.6, SSP5-8.5, and an indicative medium emission scenario (M) for CMIP7 ScenarioMIP. GMST anomalies for scenario M are computed with FaIR. (b) Joint monthly temperature–precipitation kernel density estimates for the North South America region in 2100. The region is depicted on the right. Contours for the M scenario use emulated data. Density plots for SSP1-2.6 and SSP5-8.5 use MPI-ESM1-2-LR simulations. Contour labels indicate the probability mass outside each contour.}
    \label{fig:tp-m-emulation}
\end{figure}

In this work, we make the simplifying assumption that GMST anomalies, through pattern scaling, provide a sufficient predictor for the sources of forcing relevant to project climate change. This design, recurrent in the emulator community, enables coupling of the emulator with simple climate models, thereby allowing ESM outputs to be emulated directly from global emissions pathways~\citep{beusch2021emission, mathison2025rapid}. Figure~\ref{fig:tp-m-emulation} illustrates this using an indicative medium emissions scenario representative of the forthcoming CMIP7 ScenarioMIP~\citep{chrisroadmap, van2025scenario}. GMST anomalies for this scenario are computed using FaIR~\citep[Finite amplitude Impulse Response][]{leach2021fair}, a simple climate model that estimates GMST from greenhouse gas emissions, with parameters calibrated for the MPI-ESM1-2-LR model. The emulator then projects regional temperature–precipitation distributions for the year 2100 under this scenario.

The assumption that local variables scale with GMST can, however, introduce two important sources of error~\citep{womack2026theoretical}. First, it neglects the system’s memory by assuming that responses depend only on the instantaneous GMST. While this approximation may hold under monotonously increasing forcing scenarios~\citep{giani2025origin}, it potentially introduces biases in overshoot pathways, where hysteresis effects may prevent regional climates from returning to past states~\citep{womack2025rapid}. Second, it introduces a hidden variable error by excluding the influence of forcings not directly encoded in GMST. For instance, although global precipitation is expected to decrease with reductions in GMST, it also exhibits a direct radiative cooling response to atmospheric CO\textsubscript{2} such that a temporary increase in precipitation may accompany its long-term decline~\citep{o2012energetic}. More evidently, the GMST cannot capture the impact of regional forcings, such as aerosols or ozone, which have the potential to play a significant and very localized role~\citep{williams2022strong}. Approches to overcome these deficiencies include replacing pattern scaling with an impulse-response model that accounts for system memory and allows for warming patterns to evolve over time, and expanding the conditioning signal to include emissions data alongside surface temperature, with spatially resolved maps of emission maps of regional short-lived forcings.

From a different standpoint, the instantaneous nature of this forcing also neglects the dynamical nature of the system. As a result, the emulator cannot generate temporally consistent samples, limiting its ability to study the influence of seasonal and multi-annual variability in projections. A way forward would be to adopt an autoregressive strategy \citep[e.g.][]{clark2025ace2}, conditioning the emulator not only on forcing but also on past states, to generate entire spatio-temporal sequences.

\subsection{On the practical use to support impact assessment}

Impact modeling often requires climate projections at daily resolution and spatial scales of at least $0.5^\circ$~\citep{warszawski2014inter, maraun2016bias} --- in particular for extreme events --- which is much finer than the resolution presented in this work. The current emulator may already be useful for some applications that use long-term averages, for example, in ecology~\citep{tabor2010globally, mahony2022global}, or for simple models of crop yields or water balance~\citep{ray2015climate, bock2017us}. However, we prefer to view it as a building block toward higher temporal and spatial resolution approaches. We chose to work with monthly averages to keep data volumes manageable on our limited computing resources. However, going to higher resolutions and frequencies is well within the scope of our intended work. In practice, the emulator could already be paired with statistical or dynamical downscaling techniques to provide projections on scales suitable for impact assessment \citep[e.g.][]{schillinger2025enscale}.

Another key consideration is the computational cost. The more efficient an emulator is, the more accessible it becomes --- a common claim is that emulators should run on standard personal computers. While our emulator is computationally efficient, it still requires GPU acceleration to achieve this performance. An important focus in our development has been to keep the network lightweight, but reducing computational requirements further remains an important goal for future work. At the same time, full portability may become less critical in the future. Large language models are hosted on servers and made accessible through web interfaces. Modeling centers could provide GPU-backed services to run emulators of their own ESMs, making them available to the community without requiring local hardware. Energy demands for such an approach must, however, be considered.

When evaluating an emulator, it is natural to aim for the closest possible match to its reference ESM. However, perfect agreement is not always necessary. First, given the known deficiencies of ESMs, a perfect reproduction of all aspects of their output distribution may have limited practical value for physical risk assessment. Some deviation should be tolerated where ESM projections are uncertain or biased. Second, although the relevant statistics may vary across impact models, many rely on specific quantities --- such as means, variability, or extremes --- rather than the full climate distribution. In these cases, what matters is that the emulator reproduces those relevant statistics with sufficient skill, not that it matches every feature of the ESM output. Finally, even when the ESM is reliable and the statistics of interest are well defined, we argue that small discrepancies between the emulator and the ESM are unlikely to affect downstream impact assessment, provided such deviations remain within the envelope of the ESM internal variability. In this sense, an emulator need not stringently mimic its reference ESM to be practically useful; it only needs to be accurate enough for the task it supports.

\section{Conclusion}\label{section:conclusion}

We have introduced a score-based generative emulator of monthly averaged climate model output anomalies conditioned on GMST anomalies. The emulator targets four near-surface variables of relevance for impact assessment: temperature, precipitation, relative humidity, and wind speed. It provides a computationally efficient procedure to draw samples from a distribution that approximates the joint distribution of climate model outputs. To evaluate performance, we relied on diagnostics that compared statistical features of emulator-generated data against ESM outputs under both unforced and forced regimes. In particular, we assessed whether discrepancies between emulator and ESM distributions are meaningful relative to internal variability.

Our results demonstrate that the proposed emulator is capable of producing distributions that closely match the ones from ESM outputs, showing promise to reproduce key statistical features of the reference ESM. These statistical features include higher-order moments, cross-variable correlations, and the extreme tails. Although the emulator may not be a perfect match to the ESM, it has potential to be useful to support impact assessment given the range of variability in future projections. Our analysis also reveals important failure cases, such as difficulties in representing strong seasonal distributional shifts and a tendency to overfit portions of the training data. These findings highlight both the promise of deep generative approaches for climate model output emulation and the challenges ahead. Future work should address these modeling limitations while advancing the emulator as a practical tool for impact assessment, with priorities including finer spatial and temporal resolution, reduced computational cost, and bias correction or transfer learning to bring outputs closer to observational distributions.

\section*{Acknowledgments}

This work acknowledges support by Schmidt Sciences, LLC, through the Bringing Computation to the Climate Challenge (BC3), an MIT Climate Grand Challenge Project. We also acknowledge the MIT \emph{Engaging} cluster supported by the Office of Research Computing and Data, and MIT \emph{Svante} cluster supported by the Center for Sustainability Science and Strategy for computing resources. We are grateful to Paolo Giani, Björn Lutjens, Christopher Womack, Noelle Eckley Selin, Glenn Flierl, and Claudia Tebaldi for insightful discussions and thorough feedback, which have helped shape this work. We also want to thank the three anonymous reviewers for their constructive feedbacks which have improved the quality of the manuscript. We thank Laura Battaglia for valuable discussions that greatly contributed to the development of this work.

\section*{Open research}

All data used in this work are publicly available through the World Climate Research Program (WCRP) Coupled Model Intercomparison Project 6 (CMIP6) and were retrieved through the Earth System Grid Federation interface. Code to reproduce results is published here \url{https://github.com/shahineb/climemu} (\url{https://doi.org/10.5281/zenodo.18361257}). The indicative medium emission scenario representative of the forthcoming CMIP7 ScenarioMIP is publicly available from the GitHub repository \href{https://github.com/chrisroadmap/cmip7-scenariomip}{\texttt{chrisroadmap/cmip7-scenariomip}}. We use release v1.0, published on 11 December 2024, available at \url{https://github.com/chrisroadmap/cmip7-scenariomip/releases/tag/1.0}. Pretrained model weights are published here \url{https://huggingface.co/shahineb/climemu}. Data is processed using Xarray~\citep{hoyer2017xarray}, SciPy~\citep{2020SciPy-NMeth} and NumPy~\citep{harris2020array}. Figures were made with Matplotlib~\citep{caswell2020matplotlib, hunter2007matplotlib}, Cartopy~\citep{Cartopy} and Seaborn~\citep{Waskom2021}. Models are implemented using Equinox~\citep{kidger2021equinox}, Optax~\citep{deepmind2020jax}, Diffrax~\citep{kidger2021on}, using the JAX ecosystem~\citep{jax2018github}.

\bibliographystyle{unsrtnat}
\bibliography{references}

\end{document}